# Disagreement among global cloud distributions from CALIOP, passive satellite sensors and general circulation models

V. Noel (1), H. Chepfer (2), M. Chiriaco (3), D. M. Winker (4), H. Okamoto (5), Y. Hagihara (6), G. Cesana (7), A. Lacour (2)

(1) CNRS/INSU, Laboratoire d'Aérologie/OMP, 14 avenue Edouard Belin, 31400 Toulouse, France
(2) UPMC, Laboratoire de Météorologie Dynamique, France
(3) UVSQ, Laboratoire Atmosphères, Milieux, Observations Spatiales, 11 Bd d'Alembert, 78280 Guyancourt, France
(4) NASA Langley Research Center, Hampton, VA, USA
(5) Research Institute for Applied Mechanics, Kyushu University, Fukuoka, Japan
(6) Earth Observation Research Center, Japan Aerospace Exploration Agency, Tsukuba, Japan
(7) NASA Goddard Institute for Space Studies, Department of Applied Physics and Applied Mathematics, Columbia University, NY, USA

Contact author:

Vincent Noel

Laboratoire d'Aérologie, OMP, 14 avenue Edouard Belin, 31400 Toulouse, France.

+33 5 61 33 27 55

vincent.noel@aero.obs-mip.fr

## Abstract

Cloud detection is the first step of any complex satellite-based cloud retrieval. No instrument detects all clouds, and analyses that use a given satellite climatology can only discuss a specific subset of clouds. We attempt to clarify which subsets of clouds are detected in a robust way by passive sensors, and which require active sensors. To do so, we identify where retrievals of Cloud Amounts (CAs), based on numerous sensors and algorithms, differ the most. We investigate large uncertainties, and confront retrievals from the CALIOP lidar, which detects semitransparent clouds and directly measures their vertical distribution, whatever the surface below. We document the cloud vertical distribution, opacity and seasonal variability where CAs from passive sensors disagree most.

CALIOP CAs are larger than the passive average by +0.05 (AM) and +0.07 (PM). Over land, the +0.1 average difference rises to +0.2 over the African desert, Antarctica and Greenland, where large passive disagreements are traced to unfavorable surface conditions. Over oceans, CALIOP retrievals are closer to the average of passive retrievals except over the ITCZ (+0.1). Passive CAs disagree more in tropical areas associated with large-scale subsidence, where CALIOP observes a specific multi-layer cloud population: optically thin, high-level clouds and opaque (z>7km), shallow boundary layer clouds (z<2km).

We evaluate the CA and cloud vertical distribution from 8 General Circulation Models where passive retrievals disagree and CALIOP provides new information. We find that modeled clouds are not more realistic where cloud detections from passive observations have long been robust, than where active sensors provide more reliable information.

## 1. Introduction

Passive remote sensors have been monitoring the state of the atmosphere from space for several decades (e.g. Rossow and Schiffer 1999; Norris and Slingo, 2009; Stubenrauch et al., 2013). By analyzing their measurements of radiances or reflected sunlight, we can detect clouds, and for those detected retrieve their top altitude and optical properties. Clouds are harder to detect above specific surface types: ice, snow, deserts, and more generally any surface whose temperature is hard to predict accurately. The ability to detect clouds also depends on their optical depth and dimensions, with thinnest or smallest clouds sometimes escaping detection.

An instrument detection performance defines the subset of clouds for which other properties can eventually be retrieved. A bias in detection can thus lead to biases on global cloud properties. For a properly detected cloud, retrieving its top temperature from single-channel passive measurements requires strong assumptions on cloud emissivity that are rarely realistic, and estimates of cloud top altitudes can be off by several kilometers (Sherwood et al., 2004; Holz et al, 2008; Chang et al., 2010). Multispectral passive data as in $CO_2$ slicing can provide a better accuracy (Wylie and Menzel, 1999; Wang et al., 2012) but will not work when the optical depth is too low. Considerable effort and ingenuity went into solving these problems and reducing uncertainties as much as possible, leading to reliable climatologies of cloud cover on a global scale. The error bars, however, remain significant, and interpreting the observed cloud change remains a challenge (Foster and Heidinger, 2013; Chepfer et al., 2014), while the detailed vertical cloud structure simply cannot be retrieved from passive sensors. The emergence of multi-decades passive cloud records enabled new types of observation-based studies, that attempt to understand how clouds react to changes in their atmospheric environment (e.g. Davies and Molloy, 2012; Evan and Morris, 2012; Marchand, 2013), or how clouds change under the influence of climate warming (Norris et al., 2016). As passive sensors do not detect all clouds, however, the observed changes concern a subset of all clouds, leading to a partial view of the problem. To properly understand the observed change in long-term passive time series, it is important to know precisely in which regions which cloud types are robustly observed by passive sensors and which population of clouds, in which regions, are not.

Unlike passive sensors, active sensors such as lidars and radars can generally detect even optically thin clouds with a minimal influence of the surface (Wylie et al., 2007), and document directly their distribution along the vertical dimension with high vertical resolution, by measuring the time for electromagnetic pulses to travel back and forth between the instrument and cloud particles (Di Michele et al. 2012; Mace and Zhang, 2014). Most notably, the CALIPSO (Cloud-Aerosol Lidar and Infrared Pathfinder Satellite Observation, Winker et al., 2007, 2010) and CloudSat (Stephens et al., 2002) missions have led to significant insights regarding the three-dimensional distribution of clouds. Thanks to its active sensor capability and high vertical resolution (up to 30m), the Cloud-Aerosol Lidar with Orthogonal Polarization (CALIOP), onboard CALIPSO, can detect clouds, even small cumulus, located above reflective surfaces, sunlit or not, down to the point of complete direct light scattering attenuation (optical depths ~3 to 5). This means that its measurements always accurately document whether a given atmospheric column contains clouds or not, including extremely thin cirrus clouds (down to optical depths of 0.005, Reverdy et al., 2012) and broken boundary layer shallow cumulus (Konsta et al., 2012) owing to a fine along-track spatial resolution (330m). Such measurements have been used extensively to characterize clouds at global scales, with an emphasis on the vertical distribution (e.g. Massie et al., 2010; Naud et al., 2010; Wu et al., 2011; Nair and Rajeev 2014) and evaluate their representation in models (for instance Chepfer et al., 2008; Ahlgrimm and Köhler, 2010; Cesana and Chepfer, 2012; Konsta et al., 2012; Huang et al., 2014; Nam et al., 2014; Miller et al., 2014; Wang et al., 2014).

Our objective here is to identify, using new information from active sensors, which subset of clouds has not been documented reliably during the past 20 years using passive remote sensing observations. We also investigate if these particular clouds suffer from a poor representation in climate models. After introducing the global-scale observational datasets (Sect. 2), we describe in Sect. 3 where passive sensors disagree on cloud presence. In Sect. 4, we identify and describe geographic areas where datasets from passive sensors mostly disagree among themselves and with CALIOP retrievals. To explain the differences, in Sect. 5 we use measurements from CALIOP to describe the vertical distribution of clouds in these areas and its seasonal variability. Finally, Sect. 6 investigates if large disagreements between datasets from passive sensors have had a significant effect (or not) on the representation of clouds in several General Circulation Models. Sect. 7 sums up our main results.

## 2. Observational datasets

### 2.1. Passive remote sensing

We have used cloud retrievals from the Global Energy and Water cycle Experiment (GEWEX) Cloud Assessment (Stubenrauch et al., 2013), where 13 different teams produced monthly-mean global cloud products based on measurements from spaceborne instruments, packaged in the same format and made available online[1]. Most of these datasets cover 2006 through 2009. To maximize the number of datasets used, we consider retrievals made in 2007 (as in Stubenrauch et al., 2012), when in addition to CALIOP (Sect. 2.2) nine (AM) and twelve (PM) datasets from passive sensors are available, including ISCCP. This choice excludes from the comparison TOVS (stopped in 1994) and HIRS (only 2006 and 2008).

From the GEWEX datasets, we used 1°x1° global monthly maps of Cloud Amount (CA): 0 is clear-sky, 1 indicates the bin is completely cloudy, with intermediate levels of cloudiness in-between. Passive retrievals are generally more reliable in daytime, when they can use both infrared and visible channels, than in the infrared-only nighttime. As a consequence, we considered separately observations made in daylight and eclipse conditions depending on the local observation time which, for polar-orbiting satellites, is the half-orbit Local Time of the Ascending Node (LTNA). Most datasets come from the A-Train constellation, so they were grouped around its LTNA, 1:30AM and PM. These groups include instruments with LTNA from 3h30 before the A-Train (MISR) to 1h30 after the A-Train (ISCCP). PATMOS-X retrievals at 07:30AM and PM were too distant in time and excluded. Observations in a given group therefore span 5 hours or less, a non-negligible time window whose consequences will be discussed in Sect. 3 and 4. Since the GEWEX MODIS-CERES-Terra dataset (10:30AM and PM) was also excluded due to technical issues, 6 datasets derived from passive measurements were eventually left in the 10PM-3AM "AM" group (nighttime conditions) and 9 in the 10AM-3PM "PM" group (daytime conditions), cf. Table 1. CA maps for individual datasets, averaged over 2007, are shown in the Supporting Information (Fig. S1 and S2).

---

[1] All data was obtained through http://climserv.ipsl.polytechnique.fr/gewexca.

For most datasets including CALIOP, retrievals in the night group use only observations in eclipse conditions (i.e. no sunlight), and retrievals in the day group use only observations in sunlit conditions. Transient solar noise might appear at orbit boundaries, i.e. at extreme latitudes, but was excluded. AIRS, ISCCP and PATMOS-X retrievals chose to retain the half-orbit extremities even when they are under different sunlight conditions than the equator, and therefore include measurements under different sunlight conditions in both groups.

Instead of trying to identify and explain discrepancies among instantaneous retrievals describing the same individual scene, we used monthly averages of global cloud detections. Such temporal averaging might inject additional differences among datasets. However, we are here less interested in attributing detection performances to algorithmic and/or instrument sensitivity, than in understanding how these performances translate into views of the global cloud distribution that are unique to a given dataset. Using monthly mean detections minimizes the effect of instrumental variations such as swath, spatial resolution, or viewing angle, and allows a focus on the observed cloud distribution.

## 2.2. Active Remote Sensing: CALIOP

One of our goals here is to position CALIOP's vision of clouds against the passive one. There is not, however, a single canonical CALIOP cloud dataset. We consider in our present analysis the three major CALIOP cloud datasets to provide a more realistic description of how CALIOP sees clouds. Two of these datasets were part of the GEWEX Cloud Assessment project: CALIPSO-ST (Winker et al., 2009) and CALIPSO-GOCCP (Chepfer et al., 2010). Both were derived from the same CALIOP Level 1 data and made compatible with passive retrievals by considering only the highest cloud layer detected. We will hereafter refer to those datasets as C-ST and C-GOCCP for brevity. Even though GEWEX C-ST was based on CALIPSO version 2 Level 2 data, which over-estimates the cloud cover below 4 km (Stubenrauch et al., 2012, p. 107-108), we retain this dataset in our study (Sect. 3) to enable comparisons with other GEWEX studies. For all other purposes, we used the same algorithm that produced the GEWEX C-ST dataset to derive a new dataset in GEWEX format from CALIPSO version 3 level 2 data, in which the low-cloud issue has been fixed (see Supplementary Material). The third CALIOP cloud dataset, labeled C-KU (Kyushu University), comes from processing the popular CALIPSO-based cloud masks used in CloudSat-CALIPSO cloud products (Hagihara et al., 2010), which were originally developed and tested for ship-borne cloud radar and lidar observations by Okamoto et al. (2007) in midlatitudes and by Okamoto et al. (2008) in the Tropics, these schemes were modified for application to CloudSat and CALIPSO. These products were processed with the algorithm that produced the C-GOCCP GEWEX product (Cesana et al., 2016). Variations between these three datasets expose the extent to which the Cloud Amount is not a well-posed quantity even across algorithms based on the same instrument (Marchand et al., 2010), and quantifies the variation in CA due to varying definitions and hypotheses across detection algorithms.Cloud detection algorithms often include numerous refinements and optimizations: horizontal and/or vertical averaging, detection threshold adaptation to diurnal noise fluctuations or signal attenuation, etc. Choices made when designing those refinements, with a particular science question or investigation in mind, can strongly impact the retrieved CA. We will not attempt here to quantify the differences in cloud detections among the three CALIOP datasets, nor investigate their causes; these points have already been explored in Hagihara et al., 2010, 2014; Chepfer et al., 2013; and Cesana et al., 2016. Key differences between the three datasets are summed up in Sect. S4 of the Supporting Information. Note that these differences, based on 3 datasets derived from the same instrument, are not directly comparable to the passive ISDS, which is based on 6 to 9 different sensors.

## 3. Where do satellite Cloud Amounts disagree?

Maps in Fig. 1 show the CA standard deviation across all the passive datasets, considering one map of annually averaged CA per dataset (Fig. S1 and S2). We use this standard deviation as a proxy of the disagreement in cloud detection between datasets from passive sensors, and will refer to it as inter-dataset standard deviation (IDSD) from now on. Values over oceans (in blue) and continents (in orange) were split into three categories of equivalent population, as distribution histograms show (Fig. 1, right).

Over oceans, retrievals from passive sensors generally agree quite well (IDSD < 0.12 in most areas, with CA > 0.5 as in Fig. S1). Agreement is strongest over midlatitudes and the Southern Hemisphere ocean (IDSD < 0.04). Those regions are often overcast, and feature optically thick clouds that are easy targets to detect for all sensors. Agreement is rarely so good inside the Tropics (IDSD > 0.04 overall). There, it is better along the Inter-Tropical Convergence Zone (ITCZ) and the warm pool (0.04 < IDSD < 0.06), and worst in the subsidence regions in the Atlantic and in the Pacific west of South America (IDSD > 0.06). Geographic patterns and IDSD distributions are similar in passive AM (Fig. 1, top) and PM (bottom) CAs, except over the Antarctic ice shelf. The very large disagreement (IDSD >0.15) over these areas in PM, absent in AM, comes from the misclassification of solar reflection on cloud-free ice shelves.

Over continents, agreement is worse overall (IDSD > 0.1 are frequent), especially during daytime (Fig. 1, bottom). The largest disagreements (IDSD > 0.13) are found over ice and snow surfaces in Polar regions (e.g. Greenland, Antarctica and regions partially covered by ice shelves in winter) and high elevations elsewhere (Himalayas, Andes, West U.S. and Middle to Southern Africa). IDSD are often larger in PM (Fig. 1, bottom) than in AM, even though geographical patterns look similar. Note that the ATSR and MISR datasets (only available in sunlit conditions, thus only contributing to PM IDSD) do not include CA at extreme latitudes (typically above 83°), which can affect the average passive PM IDSD there. Reasons for these variable cloud detections among datasets are multiple, and include algorithmic performance as well as sensor properties: wavelength, viewing direction, footprint size, sampling frequency and extent, etc. Explaining those differences is not the goal of the present paper, and has already been addressed in depth by the GEWEX Cloud Assessment Report (Stubenrauch et al., 2012).

The considered datasets are not based on simultaneous measurements. In the worst case, MISR vs. ISCCP, there is a five-hour difference between LTAN. Large IDSD could therefore to some extent originate in diurnal changes in the monitored cloud cover, instead of revealing variations among instruments and/or algorithm performance. Rapid changes could be due, in particular, to convective activity in daytime at low latitudes, over land and west off the coast of South America and Africa (King et al., 2013). Maps built from datasets derived from A-Train measurements only, and thus simultaneous, show weaker ISDS overall (Fig. S3 in the Supporting Information), especially during daytime over non-ice land surfaces. This is consistent with diurnal changes in cloud cover influencing the IDSD during daytime. In other cases, regions of large ISDS are similar as those described by Fig. 1. This suggests the large ISDS shown in Fig. 1 are generally due to variations in instrumental and/or algorithm performance in most areas, except perhaps over non-ice land surfaces in daylight conditions. Sect. 4 will investigate this issue further by looking at regional retrievals from individual sensors.

Maps in Fig. 2 (left column) show the absolute difference between the active and passive average CA maps. These maps were recreated on a 2°×2° grid to account for CALIOP's relatively spotty global sampling. CALIOP's coverage stops at 82° due to its orbit and narrow footprint — it cannot help constrain CA above 82° where passive spread is often large, especially over Antarctica (cf. Fig. 1). The CALIPSO average CA is globally slightly higher than the CA based on passive sensors, by +0.05 (AM) and +0.07 (PM) in absolute on average. Large differences (>+0.2 absolute CA) are mostly found over continents at high latitudes and tropical regions over oceans, consistent with regions of large IDSD (Fig. 1). The active-passive difference appears only weakly correlated with IDSD in daytime conditions, as scatterplots suggest (Fig. 2, right column). Active CAs agree best with passive CAs above midlatitude oceans, where passive retrievals also agree among themselves (IDSD<0.04, Fig. 1) thanks to the frequency of optically thick clouds in the storm tracks. Active CAs agree least with passive retrievals over oceans in AM retrievals (top), and over continents in PM retrievals (bottom), although geographical distributions appear similar in both maps.

## 4. How much do satellite Cloud Amounts disagree?

### 4.1. Over land

Maps in the top row of Fig. 3 highlight four continental areas where datasets from passive sensors disagree significantly on CA over land, based on IDSD (Sect. 2) and A-Train only simultaneous IDSD (Supporting information): Greenland, Asia, South America, and Antarctica. Agreement is weakest (IDSD > 0.15 in Fig. 1) above Greenland and Antarctica (Fig. 3, rows 2 and 5), where datasets based on visible wavelengths have trouble discriminating clouds from the ice-covered surfaces. That ice and high elevations feature prominently among those reminds us that surface conditions affect significantly the CA retrieved by passive sensors, as in e.g. Rossow, 1989. The regions here were determined by extracting large IDSD (>0.09 for AM, left column, and >0.13 for PM, right column) in particular focus areas discussed in Sect. 3. Figures below the maps show the average CA by dataset in each region, roughly ordered from North to South. Symbol orientation notes if the dataset is from the A-Train (pointing down), earlier (pointing left) or later (pointing right). Within each region we considered only grid cells in which the sampling is sufficient — i.e. latitudes above 82° are excluded from these results, as in Fig. 2. In each region, green shading shows the range of active CA, and horizontal dashed lines show the average of passive CA).

In AM conditions (left column), largest passive CAs are reported by PATMOS-X over polar regions and by AIRS over Asia and South America. Smallest CAs are reported by ISCCP and MODIS-CERES-Aqua over Asia and South America, and by AIRS over ice. In PM (right), ATSR provides the smallest CA over those regions (with POLDER the second-worse) and the largest, by far, over Asia and South America (trailed by ISCCP). As ATSR flies 3 hours earlier than the A-Train over a given region, regional cloud change between overpasses could be partly responsible for large IDSD over these land regions in daytime conditions, when convective activity is significant. However, CA from MODIS-ST-Terra and MISR, almost coincident with ATSR, are in line with the rest of the datasets, or even at the other end of the distribution (over e.g. Asia or South America). Even though MISR is known to severely underestimate CAs over land (Naud et al., 2007), this suggests that instrumental and algorithmic differences are significant in that timeframe. In any case, ATSR and POLDER significantly increase the passive IDSD over land in daytime conditions.

Regional CA from the active sensor datasets are very similar overall (i.e. the green band is thin), and generally fall on the high side, but not by the same amount everywhere. They differ more during AM than PM, probably due to the influence of solar noise on cloud detection efficiency. Antarctica is where active CAs differ most, but it is also the largest region considered here. On average, active CA are +0.15 over the average passive CA, very close or above the highest passive CA, except over South America and Asia. Over ice-covered and elevated surfaces, CAs from active sensors are generally on the high end of the range reported by datasets based on passive measurements, because the lidar is sensitive to optically very thin clouds even when located above highly reflecting surfaces. C-ST reports particularly high CAs over Antarctica, thanks to its sensitivity to tenuous ice clouds, which are frequent in that region.

Over South America, the CA drops significantly in PM compared to AM. This change is related to the significant decrease in the fraction of South America with large IDSD (compare in Fig. 1 South America day and night).

## 4.2. Over Ocean

As with the continental regions described above, we isolated four specific oceanic areas containing large passive IDSD (>0.06, cf. Fig. 1 and Sect. 3): North and South Atlantic, South Pacific, and the Arabian Sea (Fig. 4, top row). Other rows in Fig. 4 show the average CA retrieved above these regions, roughly ordered from West to East. Unlike continental regions with large IDSD, which spread from the equator to the polar caps, over oceans passive sensors disagree mostly within the Tropics.

The ordering of datasets and the range of values appear more homogeneous than over continental regions. Average passive CAs hover between 0.5 and 0.6. Among the minima in the 0.4-0.5 range we often find CAs from POLDER (PM), MODIS-CERES-Aqua (AM) and ISCCP (both). Active CAs generally span the upper half of passive CAs (i.e. from slightly above the passive average to its maxima) in AM, but move closer to the average of passive CA in PM (right column), and drop below it over the Atlantic (rows 3 and 4). Generally, C-ST provides the highest active CA and C-KU the lowest, except over the Arabian Sea in nighttime. Version 2 of C-ST, part of the GEWEX Cloud Assessment, overestimates CA in trade wind regions (Sect. S3, Supporting Information), which explains why it provides the largest CA in 6 cases out of 8. These features are quite consistent in all oceanic regions, unlike the strong variations found over land (Sect. 4.1). Also in contrast with land regions, no particular sensor stands out as providing an anomalous CA retrieval above ocean.

In daylight conditions, large CAs are consistently provided in all regions by MISR and MODIS-ST-Terra, both sensors with early LTNA (1000-1030AM) compared to the A-Train (0130PM). That their values are larger than others most probably reflect the diurnal cloud cover decrease over ocean during that time interval (King et al., 2013). That ISCCP is repeatedly found at the low end of CA distributions is also consistent with that diurnal change. However, ATSR is also a late morning sensor (1030AM) but its CA is closer to the passive average than to MISR and MODIS-ST-Terra. This supports the notion that instrument/algorithm performances still play a significant role in producing large IDSD.

In this section, we have shown that variability in CA retrievals can sometimes be traced back to large surface albedos and to actual change in cloud cover between overpasses (for daytime datasets). In other cases, the source of the variability is less clear and might be related to properties of the cloud population itself. The rest of this paper aims at characterizing the types and variability of clouds in the regions above using new information from active sensors. Despite more than 20 years of monitoring the atmosphere using passive sensors, the characteristics of clouds and their seasonal variability are still not accurately known in these regions.

## 5. Cloud vertical distributions observed from active remote sensing in target regions

In this section, we examine the cloud vertical distribution observed by the active sensor CALIOP to characterize the cloud types (including multi-layer clouds) in regions of large passive IDSD. As explained in the Supporting Information (Sect. S4), we use C-GOCCP profiles of Cloud Fraction (CF) and histograms of Height vs. Attenuated Scattering Ratio. We refer to the latter as Scattering Ratio (SR) for simplicity, as in e.g. Cesana and Chepfer, 2013. Larger SR generally mean brighter, more opaque clouds. The exception is extremely small SR < 0.001, which imply full attenuation of the lidar signal – and the presence of a cloud layer totally opaque to direct visible light.

### 5.1. Over Land

Fig. 5 shows annual SR-Height histograms (in occurrences) and monthly CF profiles (in percents) in the four land regions identified in Sect. 4.1, for AM and PM (left, right columns). As in Fig. 3, regions are arranged roughly from North to South. Clouds obviously do not go as high in the Polar regions (0-12km, top and bottom rows) as over South America or Asia (0-16km, rows 2 and 3), following the tropopause.

In Polar regions, most clouds in the 3-8km altitude range are not very bright (SR < 15) . Bright low-level clouds are present between 2 and 5 km (SR > 60). They appear to rise and become thicker geometrically during summer (JJA for Greenland, DJF for Antarctica). Meanwhile, a higher population of tenuous clouds, between 6 and 8 km of altitude, follows an opposite cycle which makes them higher and thicker in winter. Totally opaque clouds (leftmost column of SR histograms) are rare, especially in PM, and appear limited to low altitudes (1-4km) in summer time.

Over Asia, clouds of limited brightness (SR < 15) appear throughout the year in the 8-10km range. In JJA, the clouds extend upwards to reach ~16km, likely due to the edge of the Hadley cell entering the region. A strong seasonal cycle, associated with deep convective clouds in summer time, appears over South America in AM, with a significant cloud population between 12 and 16km dominating both the SR-Height histogram and the CF annual cycle. This population is absent from PM consistently with the deep convective continental diurnal cycle. Opaque clouds are fairly frequent, up to 8 km.

In AM, monthly passive IDSD (square symbols over cloud fraction profiles) are strongly correlated with high cloud presence: the thinning of the upper cloud fraction is linked to higher IDSD almost everywhere. In other words, denser high clouds lead to a better agreement between datasets derived from passive sensors. IDSD from South America PM and Antarctica AM datasets remain relatively constant month to month, and both show a remarkable constancy in their vertical cloud fractions. We find a -0.79 correlation coefficient between the integrated CF above 4 km and passive IDSD in AM (Fig. S7 in Supporting Information). In PM, the correlation keeps its sign but is much weaker (correlation coefficient -0.29).

## 5.2. Over Ocean

Fig. 6 shows SR-Height histograms and the annual cycle of Cloud Fraction profiles (as in Fig. 5), for oceanic regions identified in Sect. 4.2. Results shown use AM measurements, results using PM measurements were extremely similar.

The four regions exhibit very similar patterns: a population of high (10-16km), optically thin clouds (SR < 20) lying high above frequent thick clouds in the boundary layer (0.5-2km), with clouds going from bright (SR > 40) to fully opaque (SR < 0.001). Cloud occurrence at mid-altitudes is very low. Ignoring the defined regions and directly building a unified SR-Height histogram from monthly grid cells with large IDSD leads to the same combination of optically thin high clouds and dense low clouds. High clouds are more numerous between February and June. Over the Pacific and Atlantic oceans their occurrence decreases in JJA, with a particularly sharp drop in the Southern Hemisphere. Their occurrence increases afterwards, especially over the North Atlantic in OND. The annual cycle is different over the Arabian Sea : the maximum occurrence of high clouds is seen in JJA, following the ITCZ. The low-level clouds are more frequent between June and December over Southern Hemisphere regions, a pattern reversed in the Northern hemisphere. This could however be a side-effect of the opposite cycle driving high clouds, which when more frequent would mask the low-level clouds to lidar probing.

In AM, the correlation between passive IDSD and cloud fraction above 4km is weaker than over land and bears an opposite sign (correlation coefficient = 0.31): here, thinning of the high cloud fraction is linked to smaller passive IDSD or better agreement between datasets from passive sensors as clouds disappear to reveal the easily identified surface below. The disagreement, however, does not fall below 0.05 even when high clouds are almost nonexistent (e.g. JJA over South Atlantic Ocean), possibly due to the fragmented nature of the low clouds (Sect. 7).

## 6. Clouds simulated by CMIP5 models in target regions

In this section, we investigate if state-of-the-art climate models have been influenced by disagreements among datasets based on passive sensors, by checking whether their simulations exhibit a shared bias and/or disagree more in regions of poor passive agreement. Our hypothesis is that models should produce more diverse results where the observations are themselves most uncertain: since satellite-based retrievals are supposed to point out model biases and guide model development, the dispersion across model might be constrained by observation certainty. We evaluate the modeled CA and CF profiles against CALIOP retrievals, and examine if the models share a larger bias in the target regions compared to the global average. We consider eight models that participated in the CMIP5 experiment (Taylor et al., 2012): GDFL-CM3, IPSL-CM5B, MIROC5, bcc-csm1, CNRM-CM5, CanAM4, HadGEM2-A, and MRI-CGCM3. These models have provided synthetic CALIOP daily CA and CF profiles generated by running the COSP-lidar simulator (Chepfer et al., 2008; Bodas-Salcedo et al., 2011) over the model atmosphere. The synthetic daily CA and CF profiles are then averaged monthly and annually as gridded distributions of CA and CF profiles (as in e.g. Nam and Quaas, 2012).

Using model output at specific times, for instance CALIPSO local overpass times, instead of averaging daily, has a negligible impact on the resulting monthly cloud statistics, as the models' diurnal cycles are quite weak (e.g., Chepfer et al., 2008, Cesana and Waliser, 2016). CALIPSO synthetic datasets are currently only available for 2008 for that many models, which limits our study to that year.

## 6.1. Inter-model spread

i. Cloud Amount

From the 2008 averaged CA (2°x2° grid) for each model, we derived the standard deviation across all models within each grid cell, or the "model spread". Table 2 shows the model spread of CA, averaged over target regions (defined in Sect. 4), and over latitude bands of small IDSD for comparison. Model spread is generally larger at high latitudes over land (Antarctica, Greenland) and in the South Hemisphere over ocean. Over land and oceans, model spread in regions of large IDSD is similar to the general spread in latitude bands of small IDSD.

ii. Cloud Fraction profiles

We computed for each model the 2008 averaged Cloud Fraction profiles (2°x2°x480m grid), and from those derived the standard deviation across all models within each altitude bin and grid cell of the global domain. We then extracted all the vertical profiles of this inter-model CF spread that fell inside regions defined in Sect. 4. Figure 7 (top row) shows how these vertical profiles of inter-model spread are distributed. Over Greenland (top left) models disagree most at altitudes 5-7km ASL, and this disagreement is consistent over the entire region as the distributions are quite narrow. By comparison, over South America (column 3) models disagree most between 12-15km ASL (common in the Tropics) but the large spread shows the level of agreement varies significantly over the region. When considering regions of low IDSD within latitude bands (not shown), largest disagreements are found in the same altitude ranges but are weaker overall, hardly reaching over 0.06.

Over ocean (Fig. 7, bottom row), models generally agree very well between 3 and 9 km at Tropical latitudes, where skies are mostly cloud-free and poorly sampled. Large disagreements are generally found in the 9-16km and 0-2km ASL ranges where clouds are frequent. In the Southern Hemisphere (Pacific and South Atlantic), models seem to agree better at high altitudes (consistent over the regions) and worse at low altitudes, where model spread increases up to 0.15. Over Asia, model spread is comparatively low but uniform, never larger than 3 but staying above 1.5 in most of the troposphere (3-15km). By contrast, over the Arabian Sea models agree better at low altitudes and much worse at high altitudes.

## 6.2. Model error

i. Cloud Amount

For each model, we computed the 2008 averaged CA (2°x2° grid) and subtracted the C-GOCCP CA to derive the error in each grid cell, and finally averaged errors over all models to produce a single map of average model CA error (Table 2). Errors are always larger than the inter-model spread except in Antarctica, the Arabian Sea, and the Polar regions. This suggests the considered models are all susceptible to similar biases, which move them away from observed CAs all together. Model CA error in target regions are not significantly larger than in latitude bands of small IDSD, over land or ocean. Largest errors are found over South America and the South Atlantic, where models report 18.5% and 22.1% fewer clouds than in observations. Model error reaches 14.5% over Greenland, well above the average for polar land regions (6.5%) (dominated by Antarctica, 7.1% error).

ii. Cloud fraction profiles: Annual cycles

Monthly CF profiles, averaged from all 8 models in target regions, agree well, over land (Fig. 8, left column), with annual cycles found in AM observations (Fig. 5). Fig. S6 shows modeled and observed CF cycles together for convenience. Over Greenland and Antarctica, fluctuations in cloud altitudes are captured well; the absence of clouds below 2km ASL is due to surface elevation. The variation of high clouds at 10-14km ASL over South America, with a maximum between September and May, is particularly well represented. Over Asia the modeled rise in cloud top altitude between January and July is more linear than in actual observations. In most oceanic regions (Fig. 8, right column), model cycles compare quite well with observations (Fig. 6), even though low-level cloud occurrence is always underestimated (especially in the South Atlantic and Pacific) and too shallow (0-1.5km vs. 0-2km in observations). Over the North Atlantic Ocean (top right), the modeled annual cycle exhibits too many high clouds between May and December.

iii. Cloud fraction profiles: Model error distributions

Finally, we computed for each model the 2008 averaged CF profiles (2°x2°x480m), and subtracted observed CF from C-GOCCP, to derive the error in each grid cell. Fig. 9 shows how these errors, aggregated over models and regions, are distributed at each vertical altitude level (these errors are shown relative to the observed CF in Fig. S8 in the Supplementary Material). Over land (left column), model errors are often widely spread and reach -0.2 and +0.15, particularly in the upper troposphere. The average model error (full line), however, stays consistently in the ±0.05 range. Over oceans (right column), distributions of model errors are narrow (constrained between ±0.05) in the 2-8km ASL range, where the cloud fraction is very low. These models produce too many clouds aloft (9-13km ASL), and too few low clouds below 2km ASL, where model errors are widely spread and largely negative. These findings are consistent with Chepfer et al., 2008, Cesana et al., 2012 and Cesana and Waliser, 2016. The lack of low clouds is most significant over South Atlantic and Pacific oceans (-0.1 on average, down to -0.2), in agreement with Klein et al., 2013; Konsta et al., 2012 and Nam et al., 2012.

## 7. Summary

In this paper, we identified regions where Cloud Amounts retrieved from spaceborne passive sensors are less robust, and where active sensors are required to detect clouds. In these regions, we first quantified the disagreement in CA among retrievals from passive sensors - 9 datasets in sunlit conditions, and 6 in eclipse. We compared these results with CA and annual cycles of Cloud Fraction profiles retrieved from the spaceborne lidar CALIOP, considering three popular datasets. Finally, we quantified the inter-model spread of cloud amounts and cloud fraction profiles, considering 8 CMIP5 GCMs, and quantified the model error in CA, CF profiles and their annual cycles against C-GOCCP.

Extensive literature reports that variations in surface properties always complicate the accurate detection of clouds, and increase uncertainties in cloud detections from passive measurements (e.g. Rossow et al., 1985; Rossow, 1989; Key and Barry, 1989). The present study is, to our knowledge, the first attempt to quantify this effect across multiple global datasets in a comparative study, and to contrast them against active retrievals and model predictions. We find that unfavorable surface conditions (elevation, ice or desert) lead to anomalous extreme low and high values of CA (±0.2). Over land, the CALIOP datasets generally report significantly higher CA than most passive datasets, up to +0.2 over the African desert, Antarctica and Greenland. Over these land surfaces, once the diurnal variability of cloud cover during the measurement window has been accounted for, surface conditions appear to be largely responsible for the observed large spread in passive retrievals. In addition, the investigation of CALIOP vertical cloud distributions over land reveals that the disagreements among passive retrievals drops significantly when the high cloud fraction increases: the larger amount of high clouds hide the surface more efficiently making themselves easier to detect.

Over oceans, CALIOP retrievals are more in line with passive ones, except over the ITCZ where CALIOP CA is 0.1 larger than the passive average. The datasets from passive sensors agree quite well over ocean, in particular at midlatitudes where opaque clouds are frequent. The disagreement is the largest in tropical areas, dominated by subsidence and in the southern hemisphere (0-30°S), emphasizing the need for active sensors to better characterize subsidence boundary layer clouds. During daytime, passive ocean minima are provided by POLDER and maxima by MISR and MODIS-ST-Terra, while during nighttime ISCCP and MODIS-CE-Aqua provide the minima and MODIS-ST-Aqua and AIRS the maxima. Passive agreement is slightly better over areas associated with deep convection (ITCZ, warm pool) or in the North subtropics (0-30°N) where clouds are optically thick and overcast, a situation well represented in all datasets. In contrast with land regions, we find that over oceans the disagreement between passive datasets decreases slightly when elevated cloud fraction decreases, and the retrieval distributions are less affected by outliers and are instead rather homogeneous (±0.1 CA). Our comparisons with CALIOP vertical distributions show that large disagreements between passive datasets are linked to a specific cloud population combining a low cloud layer (0.5-2km) with high optically thin clouds (10-16km). At low altitude, fragmented cloud fields (Leahy et al., 2012, Konsta et al. 2012) lead to partially cloudy pixels in passive-derived cloud masks (Zhao and Di Girolamo, 2006), making the CA retrievals particularly sensitive to the instrument horizontal resolution (Wielicki and Parker, 1992; Pincus et al., 2012). Their combination with high cirrus, revealed by CALIOP, is probably the mixture of thin cirrus with low-level scattered shallow cumulus, created by shallow convection, which Rossow et al. (2005) found most frequent in tropical areas similar to the oceanic regions we identified in Sect. 4. This particular cloud population is not detected over land, or at least is not dominant in studied regions. In it, high clouds appear to follow a seasonal cycle, locked with the ITCZ oscillation. Their apparently smaller seasonality of low opaque clouds is harder to assess due to the presence of clouds above.

In target regions of poor agreement between datasets, models do not agree on CA and CF profiles less than in other regions. Compared to CALIOP, models generate too many high clouds and too few low clouds, especially over ocean (-22%) and in the Southern hemisphere. However, the mean model biases relative to CALIOP are not larger in target regions than elsewhere, except maybe in the trade wind regions and the upper troposphere. This confirms that the realism of CA predicted by climate models is not significantly impacted by disagreements among retrievals based on observations. This implies that the availability of spaceborne retrievals has had only a small impact on model performance, i.e. that information from such retrievals was likely not often used to constrain GCM development.

As the CALIOP lidar signal can be fully attenuated by opaque clouds, how many low clouds are reported by models and observations could be influenced by the presence of high-altitude opaque clouds. This would make model evaluation unreliable. Previous studies, however, studied the impact of high-cloud attenuation and conclude this effect is not significant enough to affect the evaluation of model biases in low-level clouds (Chepfer et al. 2008; Bastin et al., 2016). Moreover, Guzman et al. (2016) showed that full attenuation in CALIOP data happens mainly over ocean, and is due to high-altitude clouds over convective centers only. The regions considered here should not be significantly affected by this effect either. Nonetheless, we hope in the future to be able to use combined vertically-resolved CloudSat and CALIOP retrievals, which would completely document the CF profiles (outside of heavy precipitation) and their annual variability, even in clouds completely opaque to the direct visible light.

It would be interesting to investigate how other cloud properties retrieved from passive measurements and present in the GEWEX-CA fare in the identified problem areas. For instance, cloud top retrievals in the subsidence regions discussed here should be affected by the elevated tenuous cirrus detected by CALIOP. More generally, upcoming updates to the GEWEX datasets extending the period covered, adding instruments, or fixing algorithms (see e.g. Norris and Evan, 2015) will help establish the robustness of the results presented here. Future production of additional synthetic data in the context of CMIP/CFMIP exercises would let us extend our comparison beyond 2008 and evaluate the stability of our conclusions. Looking beyond CALIOP, we hope such exercises will include observations for forthcoming spaceborne lidar/radar missions such as EarthCARE (Illingworth et al., 2015) that should be soon placed in orbit.

Acknowledgments

We thank the GEWEX community for creating the various datasets part of the Cloud Assessment exercise and making them available online. Thanks to the CMIP community for producing and making available the CALIOP synthetic datasets aimed at model evaluation. We obtained and analyzed all of these datasets on the Climserv IPSL computing facilities and ICARE/CGTD data center. This research was made possible in part through financing from CNES and CNRS. Data analysis was conducted using Python and Numpy (Oliphant, 2007). Figures were created using Matplotlib (Hunter, 2007). H.Okamoto is supported by JSPS Kakenhi JP17H06139 and the Arctic Challenge for Sustainability (ArCS). H.Okamoto and Y.Hagihara are supported by the Japan Aerospace Exploration Agency for Earth-CARE Research Announcement. Gregory Cesana was supported by a CloudSat-CALIPSO RTOP at the Goddard Institute for Space Studies.

References

- Ahlgrimm, M., and Köhler, M., 2010: Evaluation of Trade Cumulus in the ECMWF Model with Observations from CALIPSO. *Mon. Wea. Rev.*, 138(8), 3071–3083. doi:10.1175/2010MWR3320.1

- Bodas-Salcedo, A. et al., 2011: COSP: Satellite simulation software for model assessment. *Bull. Am. Meteorol. Soc.* 92, 1023-1043, doi:10.1175/2011BAMS2856.1

- Cesana, G. and Chepfer, H., 2012: How well do climate models simulate cloud vertical structure? A comparison between CALIPSO-GOCCP satellite observations and CMIP5 models, *Geophys. Res. Let.*, 39(20), L20803, doi:10.1029/2012GL053153

- Cesana, G., and H. Chepfer, 2013: Evaluation of the cloud thermodynamic phase in a climate model using CALIPSO-GOCCP, J. Geophys. Res. Atmos., 118, 7922–7937, doi:10.1002/jgrd.50376

- Cesana G., H. Chepfer, D. Winker, X. Cai, B. Getzewich, H. Okamoto, Y. Hagihara, O. Jourdan, G. Mioche, V. Noel and M. Reverdy, 2016: Using in-situ airborne measurements to evaluate three cloud phase products derived from CALIPSO, *J. Geophys. Res. Atmos.,* 121, doi:10.1002/2015JD024334

- Chang, F.-L., P. Minnis, J. K. Ayers, M. J. McGill, R. Palikonda, D. A. Spangenberg, W. L. Smith Jr., and C. R. Yost, 2010: Evaluation of satellite-based upper troposphere cloud top height retrievals in multilayer cloud conditions during TC4, J. Geophys. Res., 115, D00J05, doi:10.1029/2009JD013305

- Chepfer, H., Bony, S., Winker, D., Chiriaco, M., Dufresne, J.-L., and Seze, G., 2008: Use of CALIPSO lidar observations to evaluate the cloudiness simulated by a climate model. *Geophys. Res. Let.*, 35(1), 15704. doi:10.1029/2008GL034207

- Chepfer, H., Bony, S., Winker, D., Cesana, G., Dufresne, J.-L., Minnis, P., Stubenrauch, C. and Zeng, C., 2010: The GCM-Oriented CALIPSO Cloud Product (CALIPSO-GOCCP), *J. Geophys. Res.*, 115, D00H16.

- Chepfer, H., Cesana, G., Winker, D., Getzewich, B., Vaughan, M., and Liu, Z., 2013: Comparison of Two Different Cloud Climatologies Derived from CALIOP-Attenuated Backscattered Measurements (Level 1): The CALIPSO-ST and the CALIPSO-GOCCP. *J. Atmos. Ocean. Tech.*, 30(4), 725–744. doi:10.1175/JTECH-D-12-00057.1

- Chepfer, H., V. Noel, D. Winker, M. Chiriaco, 2014: Where and when will we observe cloud changes due to climate warming? *Geophys. Res. Lett.* 41, doi:10.1002/2014GL061792

- Davies, R., and M. Molloy (2012), Global cloud height fluctuations measured by MISR on Terra from 2000 to 2010, Geophys. Res. Lett., 39, L03701, doi:10.1029/2011GL050506

- Di Michele, S., Ahlgrimm, M., Forbes, R., Kulie, M., Bennartz, R., Janisková, M. and Bauer, P., 2012: Interpreting an evaluation of the ECMWF global model with CloudSat observations: ambiguities due to

- radar reflectivity forward operator uncertainties. *Q. J. R. Meteorol. Soc.*, 138:2047–2065. doi:10.1002/qj.1936

- Evan, A. T., and J. R. Norris, 2012: On global changes in effective cloud height, Geophys. Res. Lett., 39, L19710, doi:10.1029/2012GL053171.

- Foster, M. J., & Heidinger, A., 2013: PATMOS-x: Results from a Diurnally Corrected 30-yr Satellite Cloud Climatology. *Journal of Climate*, 26(2), 414–425. http://doi.org/10.1175/JCLI-D-11-00666.1

- Guzman, R., H. Chepfer, V. Noel, T. Vaillant de Guélis, J. E. Kay, P. Raberanto, G. Cesana, M. A. Vaughan, and D. M. Winker (2017), Direct atmosphere opacity observations from CALIPSO provide new constraints on cloud-radiation interactions, *J. Geophys. Res.*, 1–20, doi:10.1002/2016JD025946.

- Hagihara, Y., H. Okamoto, and R. Yoshida, 2010: Development of a combined CloudSat-CALIPSO cloud mask to show global cloud distribution, *J. Geophys. Res.*, 115, D00H33, doi:10.1029/2009JD012344.

- Hagihara, Y., H. Okamoto, and Z. J. Luo (2014), Joint analysis of cloud top heights from CloudSat and CALIPSO: New insights into cloud top microphysics, J. Geophys. Res. Atmos., 119, 4087–4106, doi:10.1002/2013JD020919.

- Holz, R. E., S. A. Ackerman, F. W. Nagle, R. Frey, S. Dutcher, R. E. Kuehn, M. A. Vaughan, and B. Baum, 2008: Global Moderate Resolution Imaging Spectroradiometer (MODIS) cloud detection and height evaluation using CALIOP, J. Geophys. Res., 113, D00A19, doi:10.1029/2008JD009837.

- Huang, Y., Siems, S. T., Manton, M. J., and Thompson, G., 2014: An Evaluation of WRF Simulations of Clouds over the Southern Ocean with A-Train Observations. *Mon. Wea. Rev.*, 142(2), 647–667. doi:10.1175/MWR-D-13-00128.1

- Hunt, W. H., Winker, D., Vaughan, M. A., Powell, K. A., Lucker, P. L., and Weimer, C. (2009). CALIPSO lidar description and performance assessment. *J. Atmos. Ocean. Tech.*, 26, 1214–1228.

- Hunter, J. D. (2007), Matplotlib: A 2D graphics environment, *Computing in Science and Engineering*, 9 (3), 90-95.

- Illingworth, A. J., Barker, H. W., Beljaars, A., Ceccaldi, M., Chepfer, H., Cole, J., et al., 2015: The EarthCARE satellite: the next step forward in global measurements of clouds, aerosols, precipitation and radiation. *Bull. Am. Met. Soc.* doi:10.1175/BAMS-D-12-00227.1

- Key, J., and R. G. Barry (1989), Cloud cover analysis with arctic AVHRR data, *J. Geophys. Res.*, 94(D15), 18512–18535.

- Klein, S. A., Y. Zhang, M. D. Zelinka, R. Pincus, J. Boyle, and P. J. Gleckler (2013), Are climate model simulations of clouds improving? An evaluation using the ISCCP simulator, *J. Geophys. Res. Atmos.*, **118**, 1329–1342, doi:10.1002/jgrd.50141.

- King, M. D., Platnick, S., Menzel, W. P., Ackerman, S. A., & Hubanks, P. A. (2013). Spatial and Temporal Distribution of Clouds Observed by MODIS Onboard the Terra and Aqua Satellites. *IEEE Transactions on Geoscience and Remote Sensing*, *51*(7), 3826–3852. http://doi.org/10.1109/TGRS.2012.2227333

- Konsta D., H. Chepfer, JL Dufresne, 2012: A process oriented characterization of tropical oceanic clouds for climate model evaluation, based on a statistical analysis of daytime A-train observations, *Clim. Dyn.* 39 2091-2108, DOI: 10.1007/s00382-012-1533-7

- Liu, Z., Vaughan, M. A., Winker, D., Kittaka, C., Getzewich, B., Kuehn, R., et al., 2009: The CALIPSO lidar cloud and aerosol discrimination: Version 2 algorithm and initial assessment of performance. *J. Atmos. Ocean. Tech.*, 26, 1198–1213.

- Mace, G. G., and Q. Zhang, 2014: The CloudSat radar-lidar geometrical profile product (RL-GeoProf): Updates, improvements, and selected results, *J. Geophys. Res. Atmos.*, 119, 9441–9462, doi:10.1002/2013JD021374.

- Marchand, R., T. Ackerman, M. Smyth, and W. B. Rossow, 2010: A review of cloud top height and optical depth histograms from MISR, ISCCP, and MODIS, J. Geophys. Res., 115, D16206, doi:10.1029/2009JD013422.

- Marchand, R., 2013: Trends in ISCCP, MISR, and MODIS cloud-top-height and optical-depth histograms, J. Geophys. Res. Atmos., 118, 1941–1949, doi:10.1002/jgrd.50207.

- Martins, E., Noel, V. and Chepfer, H., 2011: Properties of cirrus and subvisible cirrus from nighttime Cloud-Aerosol Lidar with Orthogonal Polarization (CALIOP), related to atmospheric dynamics and water vapor, *J. Geophys. Res.*, 116(D2), doi:10.1029/2010JD014519.

- Massie, S. T., Gille, J., Craig, C., Khrosravi, R., Barnett, J., Read, W., and Winker, D., 2010: HIRDLS and CALIPSO observations of tropical cirrus. *J. Geophys. Res.*, 115, D00H11.

- Miller, S. D., C. E. Weeks, R. G. Bullock, J. M. Forsythe, P. A. Kucera, B. G. Brown, C. A. Wolff, P. T. Partain, A. S. Jones, and D. B. Johnson, 2014: Model-Evaluation Tools for Three-Dimensional Cloud Verification via Spaceborne Active Sensors. *J. Appl. Meteor. Climatol.*, 53, 2181–2195. doi: 10.1175/JAMC-D-13-0322.1

- Nair, A. K. M., and Rajeev, K. (2014). Multiyear CloudSat and CALIPSO Observations of the Dependence of Cloud Vertical Distribution on Sea Surface Temperature and Tropospheric Dynamics. *J. Clim.*, 27(2), 672–683. doi:10.1175/JCLI-D-13-00062.1

- Nam, C., S. Bony, J. L. Dufresne, and H. Chepfer (2012), The "too few, too bright" tropical low-cloud problem in CMIP5 models, Geophys. Res. Lett., 39, L21801, doi:10.1029/2012GL053421.

- Nam, C. C. W., and J. Quaas (2012), Evaluation of Clouds and Precipitation in the ECHAM5 General Circulation Model Using CALIPSO and CloudSat Satellite Data, *J. Climate*, *25*(14), 4975–4992, doi:10.1175/JCLI-D-11-00347.1.

- Nam, C. C. W., Quaas, J., Neggers, R., Siegenthaler-Le Drian, C., and Isotta, F., 2014: Evaluation of boundary layer cloud parameterizations in the ECHAM5 general circulation model using CALIPSO and CloudSat satellite data. *J. Adv. in Model. Earth Systems*, 6(2), 300–314. doi:10.1002/2013MS000277

- Naud, C., B. A. Baum, M. Pavolonis, A. Heidinger, R. Frey, H. Zhang, 2007: Comparing of MISR and MODIS Cloud-top heights in the presence of cloud overlap. *Remote Sensing of Environment* vol. 107, 200–210.

- Naud, C. M., Del Genio, A. D., Bauer, M., and Kovari, W., 2010: Cloud Vertical Distribution across Warm and Cold Fronts in CloudSat–CALIPSO Data and a General Circulation Model. *J. Clim.* 23(12), 3397–3415. doi:10.1175/2010JCLI3282.1

- Noel, V., and Chepfer, H., 2010: A global view of horizontally oriented crystals in ice clouds from Cloud-Aerosol Lidar and Infrared Pathfinder Satellite Observation (CALIPSO). *J. Geophys. Res.* 115(6). doi:10.1029/2009JD012365

- Norris, J. R., and A. Slingo (2009), Trends in observed cloudiness and Earth's radiation budget: What do we not know and what do we need to know?, in Clouds in the Perturbed Climate System, edited by J. Heintzenberg and R. J. Charlson, pp. 17–36, MIT Press, Cambridge, Mass

- Norris, J. R., & Evan, A. T., 2015: Empirical Removal of Artifacts from the ISCCP and PATMOS-x Satellite Cloud Records. *Journal of Atmospheric and Oceanic Technology*, *32*(4), 691–702. doi:10.1175/JTECH-D-14-00058.1

- Okamoto, H., et al. (2007), Vertical cloud structure observed from shipborne radar and lidar: Midlatitude case study during the MR01/K02 cruise of the research vessel Mirai, J. Geophys. Res., 112, D08216, doi:10.1029/2006JD007628.

- Okamoto, H., T. Nishizawa, T. Takemura, K. Sato, H. Kumagai, Y. Ohno, N. Sugimoto, A. Shimizu, I. Matsui, and T. Nakajima (2008) Vertical cloud properties in the tropical western Pacific Ocean: Validation of the CCSR/NIES/FRCGC GCM by shipborne radar and lidar, J. Geophys. Res., 113, D24213, doi:10.1029/2008JD009812.

- Okamoto, H., Sato, K., and Hagihara, Y., 2010: Global analysis of ice microphysics from CloudSat and CALIPSO: Incorporation of specular reflection in lidar signals. *J. Geophys. Res.*, 115(D22). doi:10.1029/2009jd013383

- Oliphant, T. E. (2007), Python for Scientific Computing, *Computing in Science and Engineering*, 9 (3), 10-20.

- Pincus, R., S. Platnick, S. A. Ackerman, R. S. Hemler, and R. J. Patrick Hofmann (2012), Reconciling Simulated and Observed Views of Clouds: MODIS, ISCCP, and the Limits of Instrument Simulators, *J. Climate*, *25*(13), 4699–4720, doi:10.1175/JCLI-D-11-00267.1.

- Reverdy, M., Noel, V., Chepfer, H. and Legras, B., 2012: On the origin of subvisible cirrus clouds in the tropical upper troposphere, *Atmos. Chem. Phys.*, 12(24), 12081–12101, doi:10.5194/acp-12-12081-2012.

- Reverdy M., H. Chepfer, D. Donovan, V. Noel, G. Cesana, C. Hoareau, M. Chiriaco, S. Bastin, 2015: An EarthCARE/ATLID simulator to evaluate cloud description in climate models. Journal of Geophysical Research – Atmospheres, doi: 10.1002/2015JD023919

- Rossow, W B, F Mosher, E Kinsella, A Arking, M Desbois, E Harrison, P Minnis, et al. 1985. ISCCP Cloud Algorithm Intercomparison. *Journal of Climate* 24 (9): 877–903.

- Rossow, W. B. (1989), Measuring Cloud Properties from Space: A Review, *J. Climate*, *2*(3), 201–213.

- Rossow, W.B., R.A. Schiffer, 1999: Advances in understanding clouds from ISCCP. *Bull. Amer. Meteor. Soc.,* **80**, 2261-2287

- Rossow, W. B., G. Tselioudis, A. Polak, and C. Jakob (2005), Tropical climate described as a distribution of weather states indicated by distinct mesoscale cloud property mixtures, *Geophys. Res. Lett.*, *32*(21), L21812–4.

- Sherwood, S., Minnis, P., & McGill, M. (2004). Deep convective cloud-top heights and their thermodynamic control during CRYSTAL-FACE. *Journal of Geophysical Research*, *109*, D20119.

- Stephens, G. L., Vane, D. G., Boain, R. J., Mace, G. G., Sassen, K., Wang, Z., et al., 2002: The Cloudsat Mission and the A-Train. *Bull. Am. Met. Soc.*, 83(1), 1771–1790. doi:10.1175/BAMS-83-12-1771

- Stubenrauch, C., W. Rossow and S. Kinne, 2012: Assessment of Cloud Data Sets from Satellites. A project of the World Climate Research Program Global Energy and Water Cycle Experiment (GEWEX) Radiation Panel. WCRP Report No. 23/2012. Available online at http://www.wcrp-climate.org/documents/GEWEX_Cloud_Assessment_2012.pdf

- Stubenrauch C. J., W. B. Rossow, S. Kinne, S. Ackerman, G. Cesana, H. Chepfer, L. Di Girolamo, B. Getzewich, A. Guignard, A. Heidinger, B. C. Maddux, W. P. Menzel, P. Minnis, C. Pearl, S. Platnick, C. Poulsen, J. Riedi, S. Sun-Mack, A. Walther, D. Winker, S. Zeng, G. Zhao, 2013: Assessment of Global Cloud Datasets from Satellites: Project and database initiated by the GEWEX radiation panel. *Bull. Am. Met. Soc.* doi: 10.1175/BAMS-D-12-00117

- Taylor, K. E., Stouffer, R. J., & Meehl, G. A., 2012: An overview of CMIP5 and the experiment design. *Bull. Am. Met. Soc.* 93, 485–498. 10.1175/BAMS-D-11-00094.1

- Wang, C., S. Ding, P. Yang, B. Baum, and A. E. Dessler, 2012: A new approach to retrieving cirrus cloud height with a combination of MODIS 1.24- and 1.38-mm channels, Geophys. Res. Lett., 39, L24806, doi:10.1029/2012GL053854

- Wang, F., X. Xin, Z. Wang, Y. Cheng, J. Zhang, S. Yang, 2014: Evaluation of cloud vertical structure simulated by recent BCC_AGCM versions through comparison with CALIPSO-GOCCP data. *Adv. Atmos. Sci.* 31 (3), 721-733, 2014. doi:10.1007/s00376-013-3099-7

- Wielicki, B. A., and L. Parker, 1992: On the determination of cloud cover from satellite sensors: The effect of sensor spatial resolution, *J. Geophys. Res.*, *97* (D12), 12799–12823, doi:10.1029/92JD01061.

- Winker, D. M., W. H. Hunt, and M. J. McGill, 2007: Initial performance assessment of CALIOP, Geophys. Res. Lett., 34, L19803, doi:10.1029/2007GL030135.

- Winker, D. M., Vaughan, M. A., Omar, A., Hu, Y., Powell, K. A., Liu, Z., Hunt W. H., and S. A. Young, 2009: Overview of the CALIPSO mission and CALIOP data processing algorithms. *J. Atmos. Ocean. Tech.*, 26, 2310–2323.

- Winker, D. M., Pelon, J., Coakley, J. A., Ackerman, S. A., Charlson, R. J., Colarco, P. R., Flamant, P., Fu, Q., Hoff, R. M., Kittaka, C., Kubar, T. L., Le Treut, H., McCormick, M. P., Mégie, G., Poole, L. R., Powell, K. A., Trepte, C., Vaughan, M. A. and Wielicki, B. A., 2010: The CALIPSO mission: A Global 3D view of Aerosols and Cloud, *Bull. Am. Met. Soc.*, 1211–1229, doi:10.1175/2010BAMS3009.1.

- Wu, D., Hu, Y., McCormick, M. P., and Yan, F., 2011: Global cloud-layer distribution statistics from 1 year CALIPSO lidar observations. *Int. J. Rem. Sens.*, 32(5), 1269–1288. doi:10.1080/01431160903530821

- Wylie, D. P., and W. P. Menzel (1999), Eight years of high cloud statistics using HIRS, J. Clim., 12, 170–184, doi:10.1175/1520-0442-12.1.170

- Wylie, D., Eloranta, E., Spinhirne, J. D., and Palm, S. P., 2007: A Comparison of Cloud Cover Statistics from the GLAS Lidar with HIRS. *J. Clim.*, 20(19), 4968–4981. doi:10.1175/JCLI4269.1

- Zhao, G., and L. Di Girolamo (2006), Cloud fraction errors for trade wind cumuli from EOS-Terra instruments, *Geophys. Res. Lett.*, *33*(20), L20802–5, doi:10.1029/2006GL027088.

**Tables**

Table 1: Grouping of spaceborne cloud datasets from the GEWEX Cloud Assessment used in the present study.

Table 2: Spread and error vs. CALIOP of model Cloud Amounts. Spread is defined as the mean, inside regions or latitude bands, of gridded standard deviation across models. Error is defined as the mean, inside regions or latitude bands, of the gridded error averaged across models.

| Group | LTAN | Sensors |
|---|---|---|
| 0130AM±5H<br>9 datasets<br>6 passive<br>3 active | 1030PM | MODIS-ST-Terra |
| | 0130AM | AIRS<br>MODIS-CE-Aqua<br>MODIS-ST-Aqua<br>PATMOS-X<br>CALIPSO-ST<br>CALIPSO-GOCCP<br>CALIPSO-KU |
| | 0300AM | ISCCP |
| 0130PM±5H<br>12 sets<br>9 passive<br>3 active | 1000AM | MISR |
| | 1030AM | MODIS-ST-Terra<br>ATSR |
| | 0130PM | AIRS<br>MODIS-CE-Aqua<br>MODIS-ST-Aqua<br>PATMOS-X<br>POLDER<br>CALIPSO-ST<br>CALIPSO-GOCCP<br>CALIPSO-KU |
| | 0300PM | ISCCP |

Table 1: Grouping of spaceborne cloud datasets from the GEWEX Cloud Assessment used in the present study.

|  |  | spread | error |  | spread | error |
|---|---|---|---|---|---|---|
| Land | Greenland | 0.10 | -0.15 | Tropics | 0.09 | -0.12 |
|  | Asia | 0.08 | -0.11 | Midlat | 0.09 | -0.14 |
|  | South America | 0.09 | -0.18 | Polar | 0.13 | -0.06 |
|  | Antarctica | 0.14 | -0.07 |  |  |  |
| Ocean | Arabian Sea | 0.09 | -0.06 | Tropics | 0.09 | -0.12 |
|  | N Atlantic | 0.09 | -0.11 | Midlat | 0.09 | -0.15 |
|  | S Atlantic | 0.12 | -0.22 | Polar | 0.11 | -0.07 |
|  | S Pacific | 0.12 | -0.15 |  |  |  |

Table 2: Spread and error vs. CALIOP of model Cloud Amounts. Spread is defined as the mean, inside regions or latitude bands, of gridded standard deviation across models. Error is defined as the mean, inside regions or latitude bands, of the gridded error averaged across models. Error is always negative: models always lack clouds compared to observations.

**Figures**

Fig. 1: Global maps and distributions of Passive Inter-Dataset Standard Deviation (IDSD), considering a single annual cloud amount map for each dataset ±3h around 0130AM UTC (top) and 0130PM UTC (bottom) over 2007. Blue is for ocean and orange for land.

Fig. 2: Difference between the average passive CA, considering a single annual map per each passive sensor (as in Fig. 1), and the average active CA, considering an annual map per each CALIOP datasets (apart from C-ST v.2), ±5h around 0130AM UTC (top) and 0130PM UTC (bottom) over 2007. Values are positive (red) where CALIOP retrieves more clouds than the passive sensors on average.

Fig. 3. First row: Four continental areas with large passive IDSD in AM (left column) and PM (right column). Rows 2-5: average CA by sensor within each area, from North to South: Greenland, Asia, South America, Antarctica. The green area shows the variability of CA retrieved from CALIOP. The dashed line shows the regional average CA retrieved from available passive sensors. ISCCP retrievals are red. Sensor and platform names are abbreviated as follows: C=CALIPSO, MS=MODIS-ST, MC=MODIS-CE, A=AQUA, and T=TERRA. Triangles point down for datasets with 0130 LTAN as the A-Train, left for earlier datasets (MISR, MODIS-ST-Terra, ATSR) and right for later datasets (ISCCP).

Fig. 4. First row: Four oceanic areas with large passive IDSD in AM (left column) and PM (right column). Rows 2-5: average CA by sensor within each area: North Atlantic, South Atlantic, Pacific, and Arabic Peninsula/Indian Ocean. Colors and sensor names as in Fig. 3.

Fig. 5: Vertical profiles of Attenuated Scattering Ratio histograms (SR, columns 1, 3) and annual cycle of Cloud Fraction profiles (CF, columns 2, 4) over land regions as in Fig. 3a. Hatching at low altitudes shows where the CALIOP sampling covers less than 50% of the area, due to high surface elevation. SR and CF were derived from C-GOCCP data over 2007. SR Counts are in thousands. Passive IDSD retrieved from monthly maps over specific regions are plotted over CF profiles.

Fig. 6: As Fig 5, over oceanic regions as in Fig. 4a. Regions were defined using AM IDSD (results using regions defined from PM IDSD were extremely similar and are not shown).

Fig. 7: distributions of inter-model CF spread at each altitude level, over land regions as in Fig. 3a (top row), and over ocean regions as in Fig. 4a (bottom row). Model spread at a given altitude is the standard deviation between annually averaged cloud amount grids (2°x2°x480m) from 8 models participating in the CMIP5 experiment (see text). The line shows at a given altitude the spread amongst models averaged within the region. Shown results use regions from AM IDSD, regions defined from PM IDSD lead to similar results (not shown).

Fig. 8: Annual cycle of synthetic CF profiles, averaged over all models (as in Fig. 7) within land (left column) and oceanic (right column) regions as in Fig. 3a and 4a.

Fig. 9: Distributions of error in CF profiles for models as in Fig. 7 compared to CALIPSO observations in land (left column) and oceanic (right column) regions as in Fig. 3a and 4a. The full line shows the average models error at each altitude.

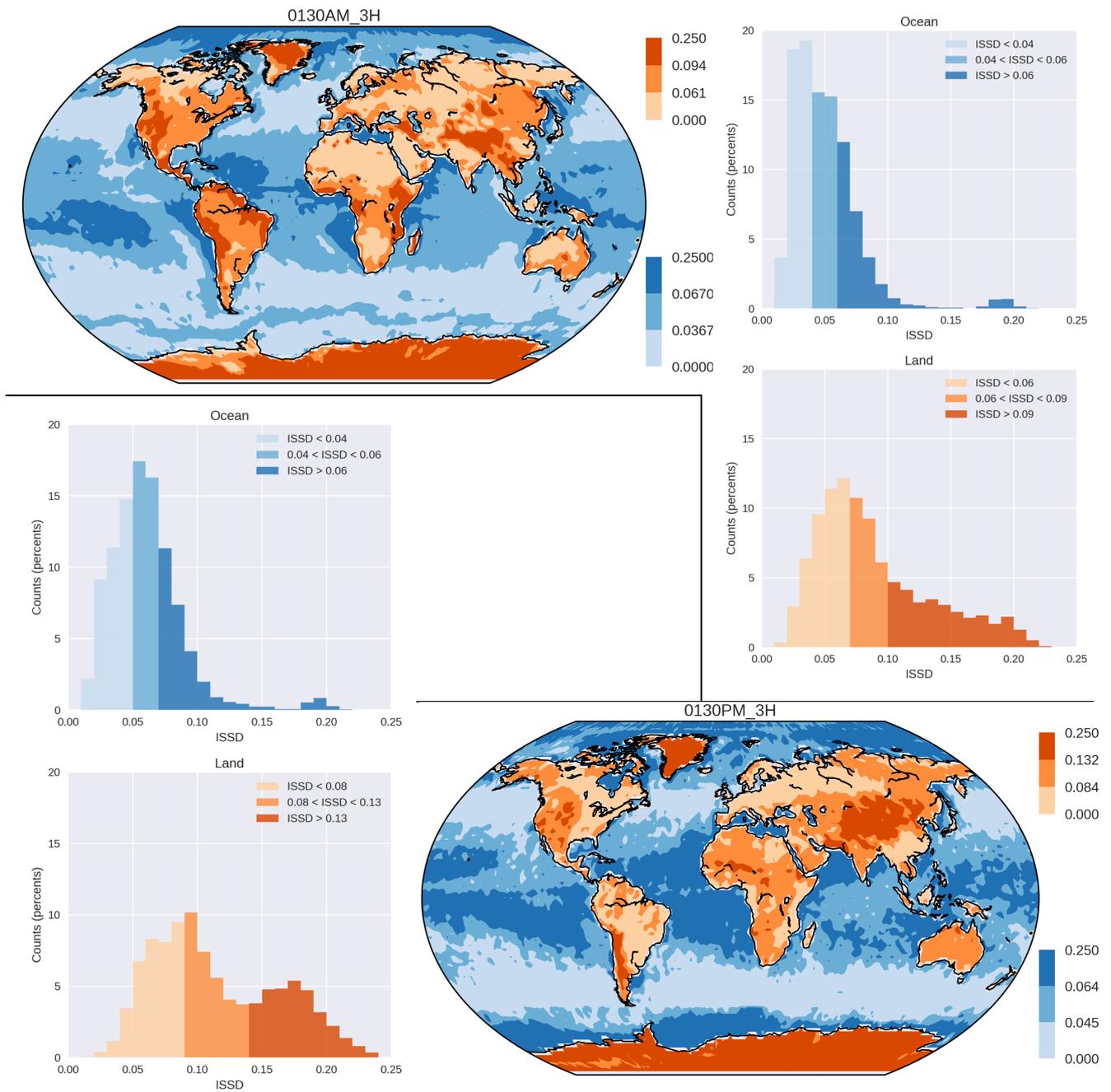

Fig. 1: Global maps and distributions of Passive Inter-Dataset Standard Deviation (IDSD), considering a single annual cloud amount map for each dataset ±3h around 0130AM UTC (top) and 0130PM UTC (bottom) over 2007. Blue is for ocean and orange for land.

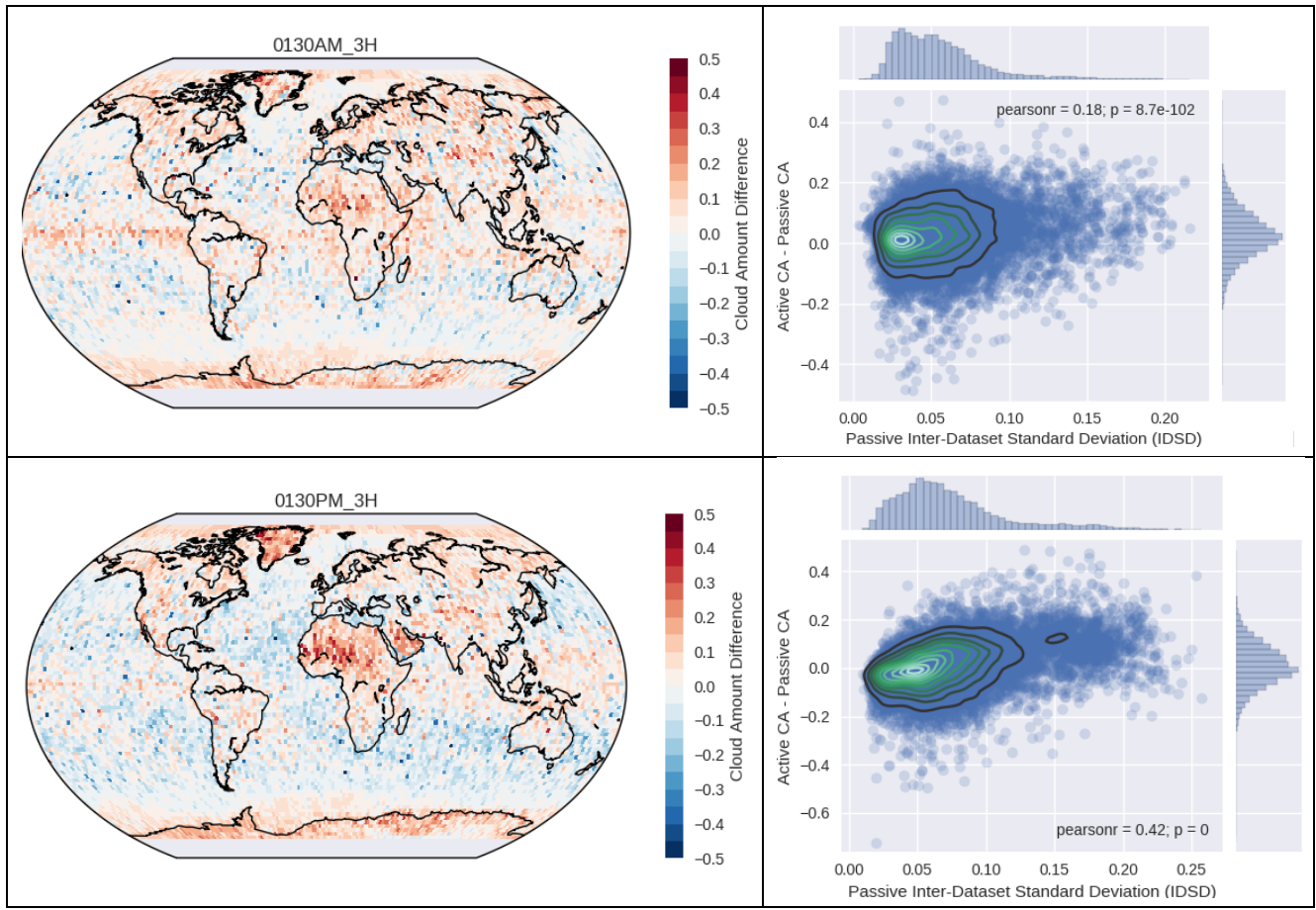

Fig. 2: left column: Difference between the average passive CA, considering a single annual map per each passive sensor (as in Fig. 1), and the average active CA, considering an annual map per each CALIOP datasets (apart from C-ST v.2), ±5h around 0130AM UTC (top) and 0130PM UTC (bottom) over 2007. Values are positive (red) where CALIOP retrieves more clouds than the passive sensors on average. Right column: distributions of the differences from the left column vs. the Passive IDSD shown in Fig. 1.

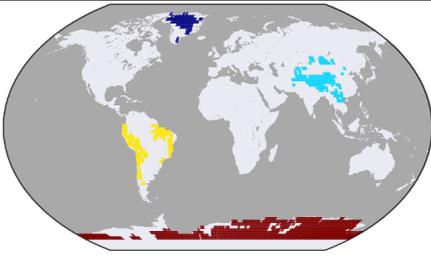
AM
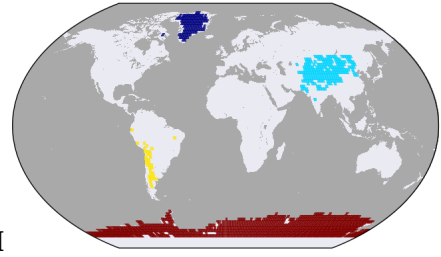
PM
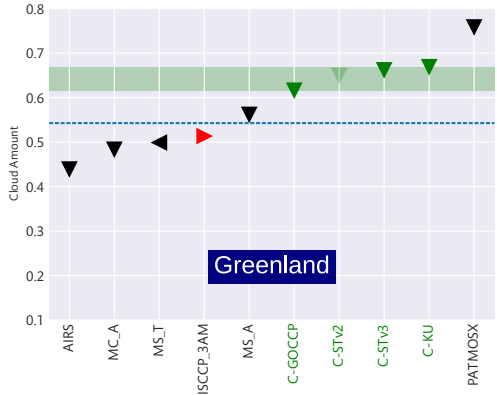
Greenland
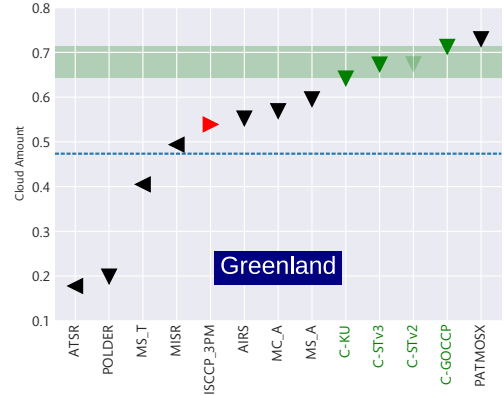
Greenland
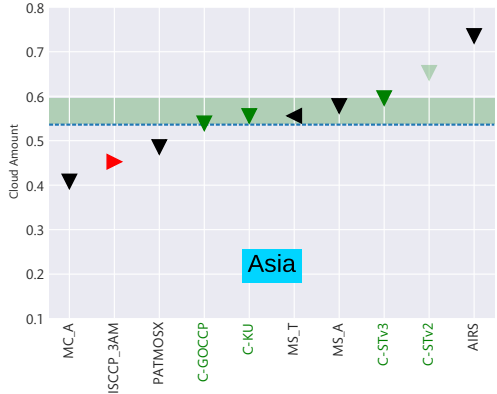
Asia
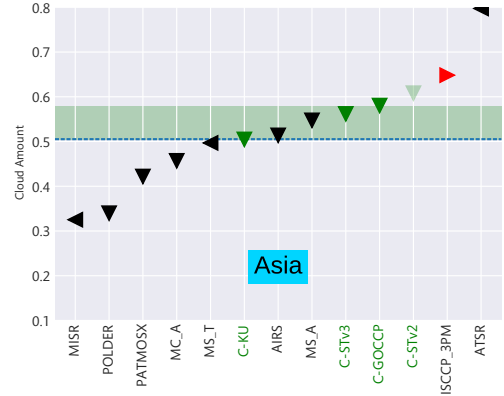
Asia
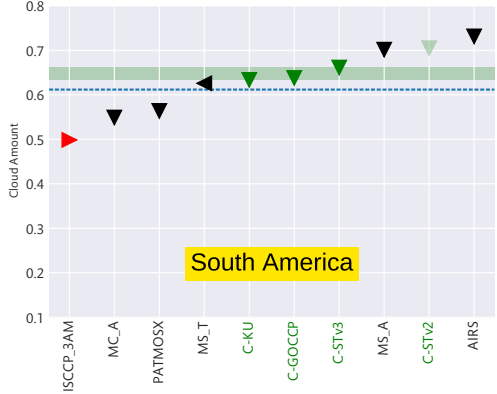
South America
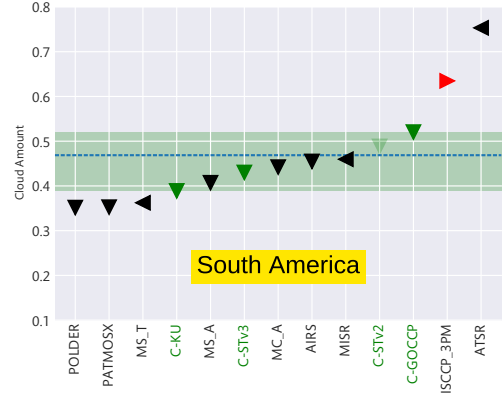
South America
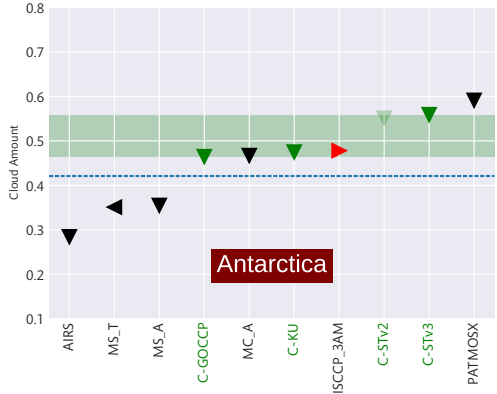
Antarctica
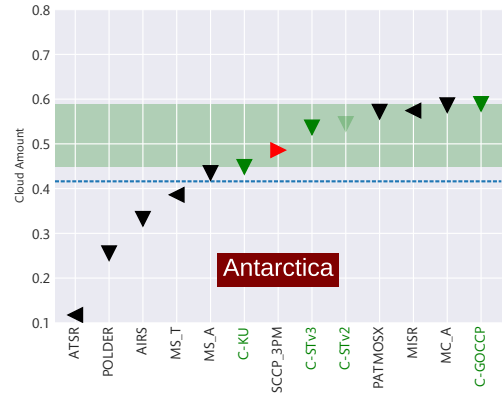
Antarctica

Fig. 3. First row: Four continental areas with large passive IDSD in AM (left column) and PM (right column). Rows 2-5: average CA by sensor within each area, from North to South: Greenland, Asia, South America, Antarctica. The green area shows the variability of CA retrieved from CALIOP. The dashed line shows the regional average CA retrieved from available passive sensors. ISCCP retrievals are red. Sensor and platform names are abbreviated as follows: C=CALIPSO, MS=MODIS-ST, MC=MODIS-CE, A=AQUA, and T=TERRA. Triangles point down for datasets with 0130 LTAN as the A-Train, left for earlier datasets (MISR and ATSR in daytime, MODIS-ST-Terra) and right for later datasets (ISCCP).

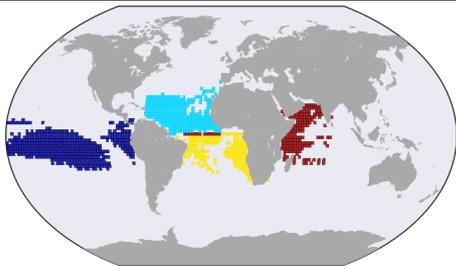 AM 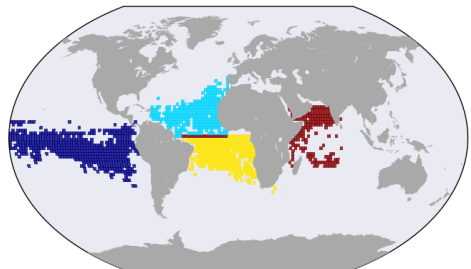 PM

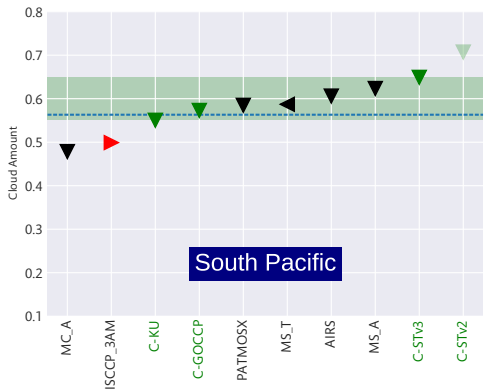
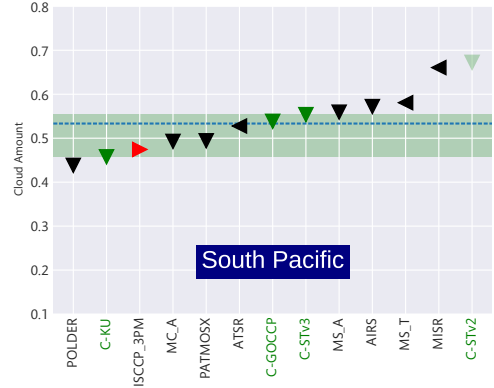

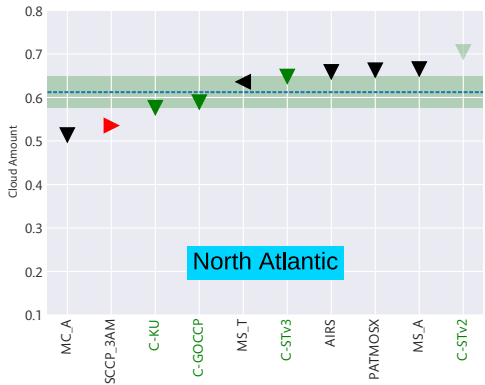
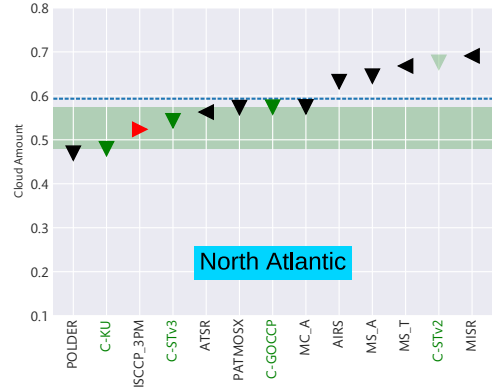

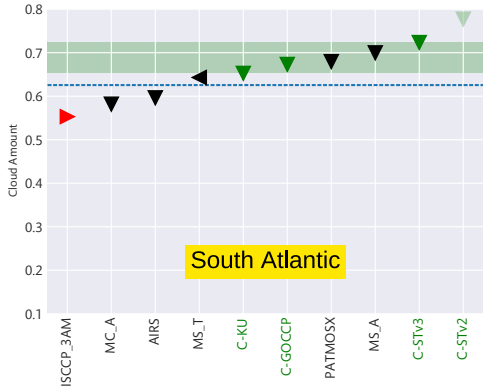
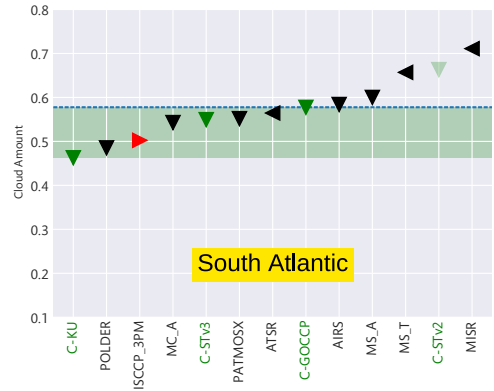

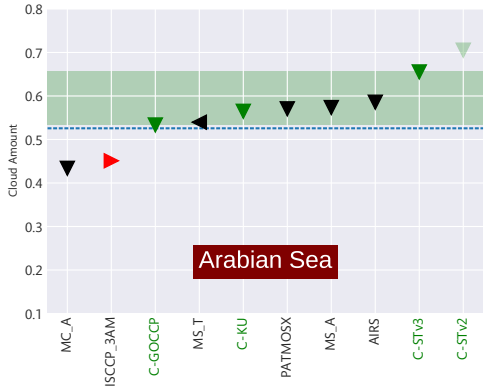
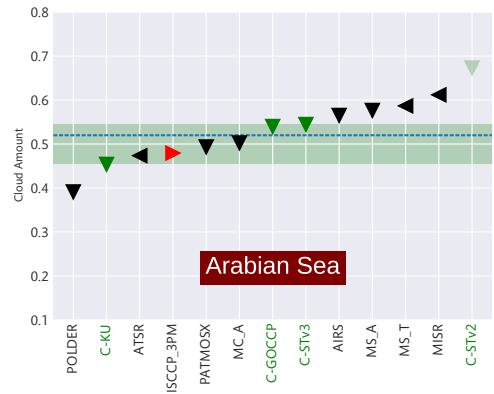

Fig. 4. First row: Four oceanic areas with large passive IDSD in AM (left column) and PM (right column). Rows 2-5: average CA by sensor within each area: North Atlantic, South Atlantic, Pacific, and Arabic Peninsula/Indian Ocean. Colors and sensor names as in Fig. 3.

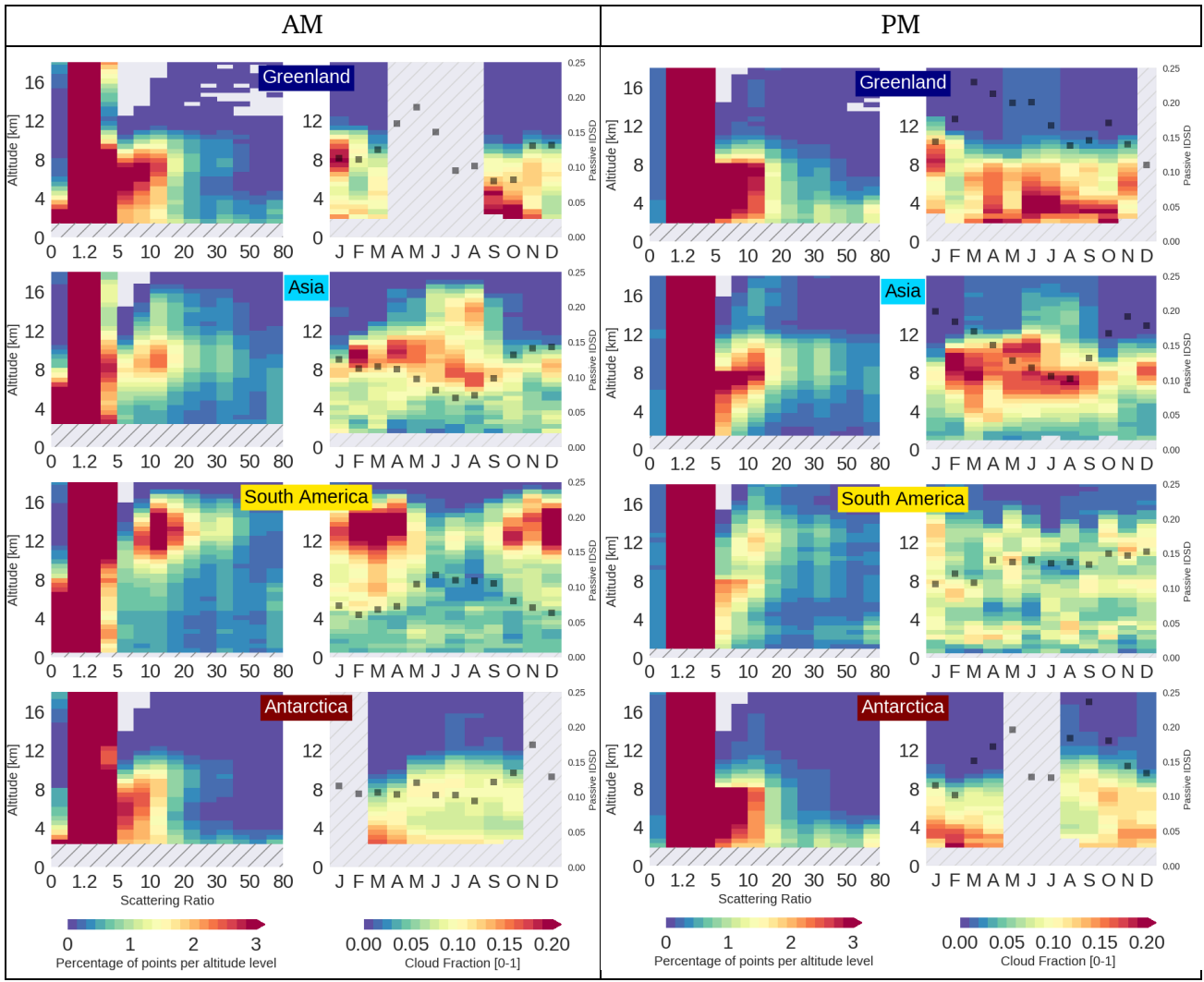

Fig. 5: Vertical profiles of Attenuated Scattering Ratio histograms (SR, columns 1, 3) and annual cycle of Cloud Fraction profiles (CF, columns 2, 4) over land regions as in Fig. 3a. Hatching at low altitudes shows where the CALIOP sampling covers less than 50% of the area, due to high surface elevation. SR and CF were derived from C-GOCCP data over 2007. SR Counts are in thousands. Passive IDSD retrieved from monthly maps over specific regions are plotted over CF profiles.

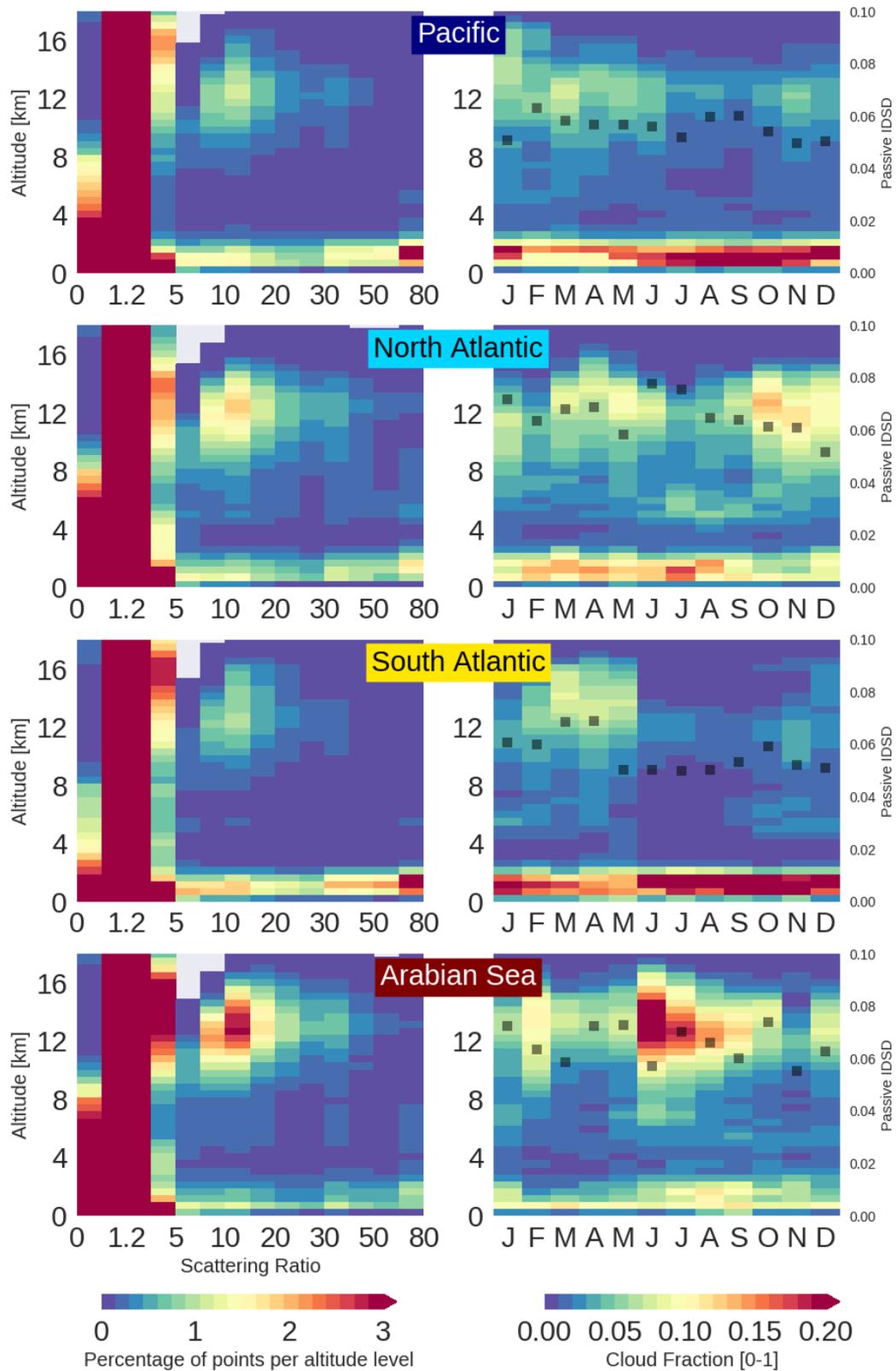

Fig. 6: As Fig 5, over oceanic regions as in Fig. 4a. Regions were defined using AM IDSD (results using regions defined from PM IDSD were extremely similar and are not shown).

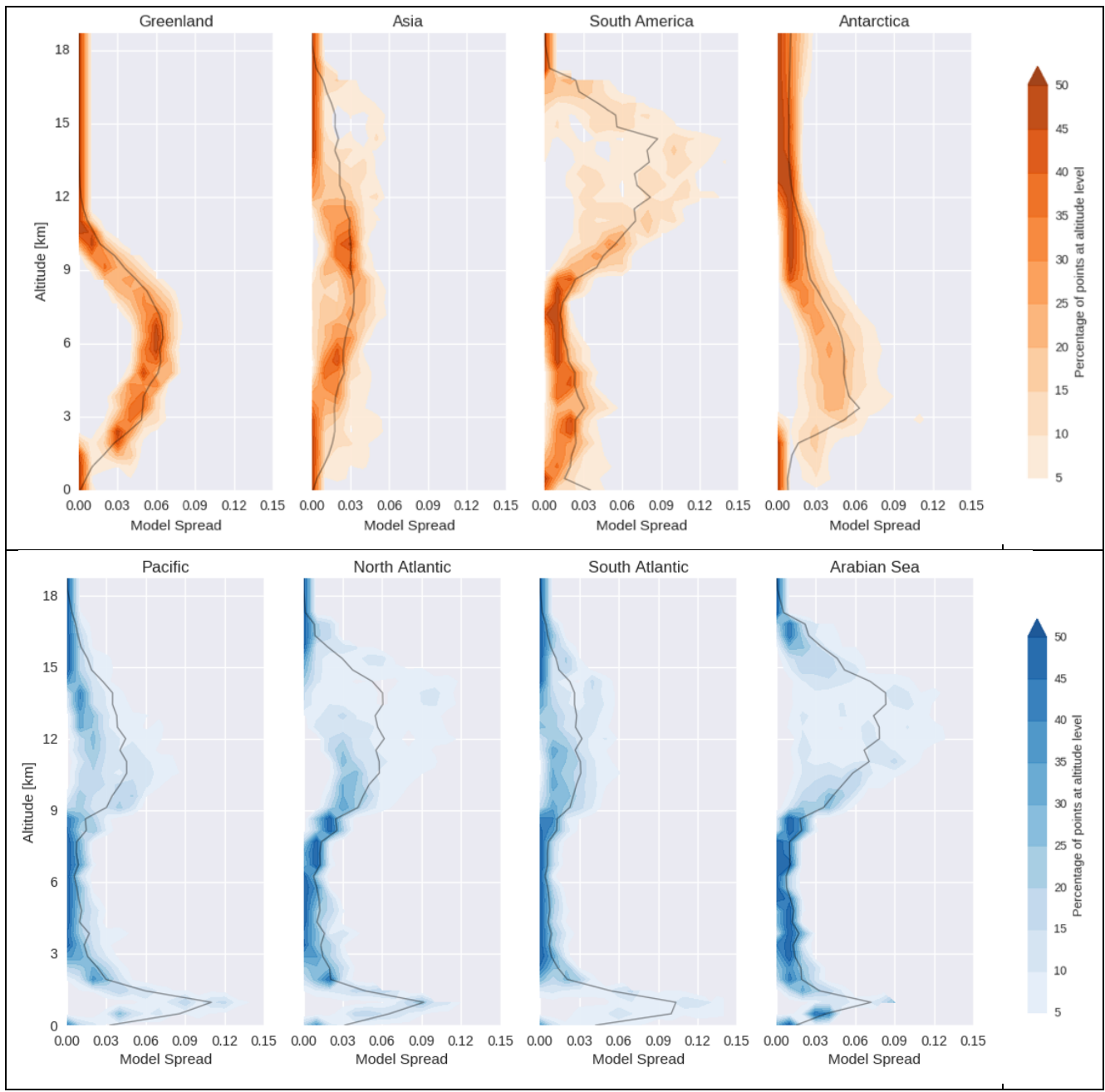

Fig. 7: distributions of inter-model CF spread at each altitude level, over land regions as in Fig. 3a (top row), and over ocean regions as in Fig. 4a (bottom row). Model spread at a given altitude is the standard deviation between annually averaged cloud amount grids (2°x2°x480m) from 8 models participating in the CMIP5 experiment (see text). The line shows at a given altitude the spread amongst models averaged within the region. Shown results use regions from AM IDSD, regions defined from PM IDSD lead to similar results (not shown).

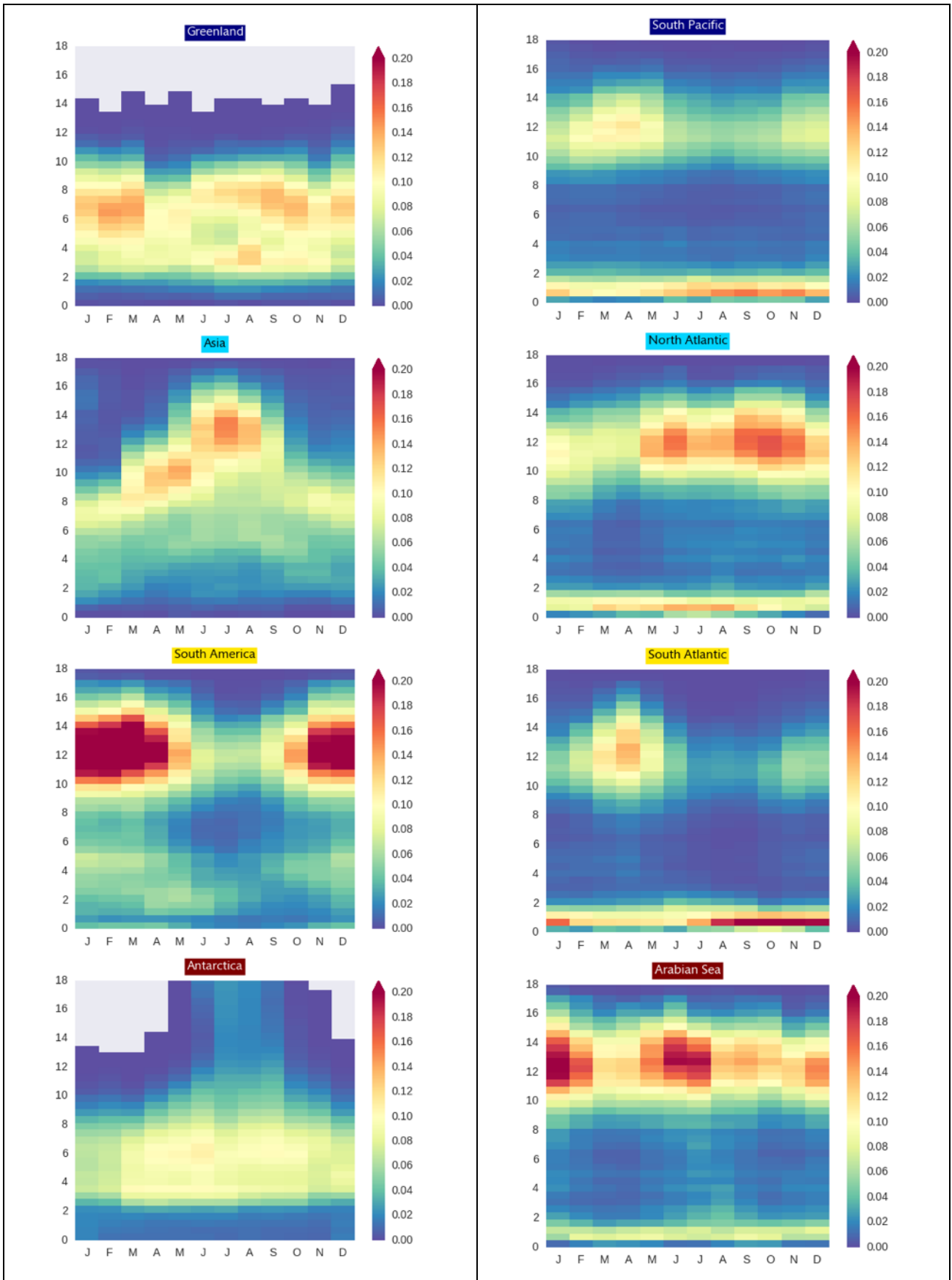

Fig. 8: Annual cycle of synthetic CF profiles, averaged over all models (as in Fig. 7) over land (left column) and ocean (right) as in Fig. 3a and 4a.

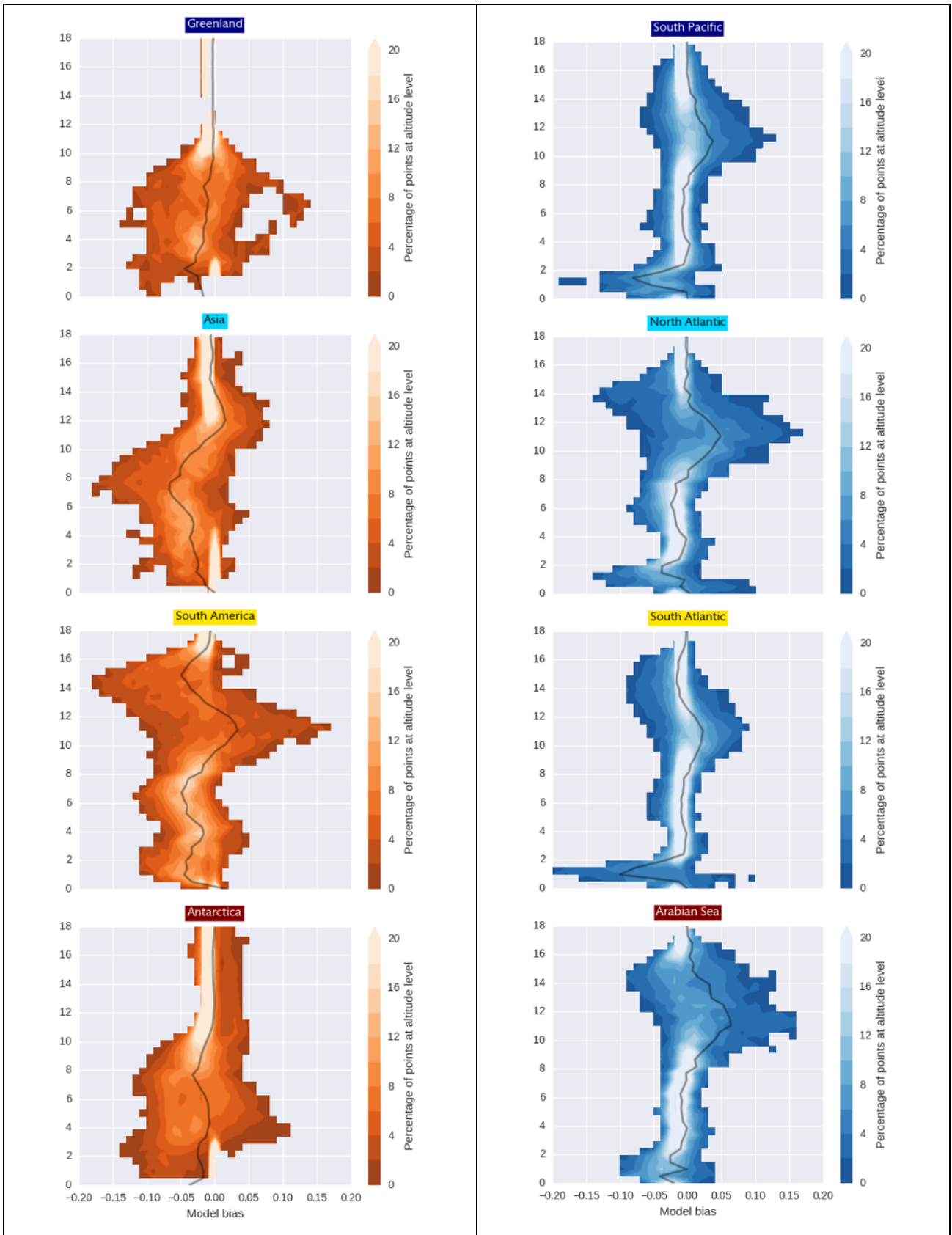

Fig. 9: Distributions of error in CF profiles for models as in Fig. 7 compared to CALIPSO observations over land (left) and ocean (right) as in Fig. 3a and 4a. The full line shows the average models error at each altitude.

Supporting Information for

# Disagreement among global cloud distributions from CALIOP, passive satellite sensors and general circulation models

V. Noel (1), H. Chepfer (2), M. Chiriaco (3), D. M. Winker (4), A. Lacour (2)

1. CNRS/INSU, Laboratoire d'Aérologie, 14 avenue Edouard Belin, 31400 Toulouse, France

2. UPMC, Laboratoire de Métérologie Dynamique, Ecole Polytechnique, 91128 Palaiseau, France

3. UVSQ, Laboratoire Atmosphères, Milieux, Observations Spatiales, 11 bd d'Alembert, 78280 Guyancourt, France

4. NASA Langley Research Center, Hampton, VA, USA

**Contents of this file**

Text S1 to S4
Figures S1 to S8

**Introduction**

This supporting information file contains a description of average Cloud Amount (CA) maps per sensor used in the main document (S1), a description of standard deviation among passive datasets with 0130 local overpass times only (S2), a comparison of the four CA datasets derived from CALIPSO observations (S3), a description of how Scattering Ratio histograms were build and should be interpreted (S4), and scatterplots of active cloud fraction integrated above 4km vs. passive IDSD (Fig. S7).

**S1, S2. Average CA maps from all sensors.**

The 2007 average CA maps for individual passive and active datasets, on which maps in Fig. 1 and Fig. 2 are based, are shown in Fig. S1 (AM) and S2 (PM). Zonal mean CA for 2007 from the four considered CALIOP datasets (Sect. 2.2 and S2) are shown in Fig. S3.

**S3. Passive Inter-Dataset Standard Deviation for 0130AM and 0130PM overpasses only**

Maps in Fig. S3 show the Inter-Dataset Standard Deviation when considering only datasets that rely on instruments with overpass local times of 0130AM and 0130PM. This only includes instruments from the A-Train.

**S4. CA datasets retrieved from CALIOP data.**

Fig. S4 compares zonally averaged cloud amounts from the four CALIOP datasets considered in the main paper. C-GOCCP cloud detection can miss optically very thin clouds (typically optical depths below 0.07, Chepfer et al., 2013) that are frequent in the Tropics (Martins et al., 2011). On the other hand, C-ST can potentially overestimate the occurrence of very thin cirrus clouds due to the averaging used to detect optically thin layers (Stubenrauch et al. 2012; Hagihara et al., 2014; Chepfer et al., 2013). C-KU CAs are lowest by 5-10% almost everywhere, even lower than C-GOCCP, especially during daytime (Fig. S4, bottom). Finally, C-GOCCP alone does not attempt to weed out aerosol layers from its cloud dataset, meaning extremely high aerosol contents could bias CA high over specific regions, for instance deserts such as the Sahara (Fig. S4). This is however rare (Reverdy et al., 2015).

C-ST v.2, used in the GEWEX Cloud Assessment, is known to over-represent low-level boundary layer clouds and thus to overestimate CA in specific regions (Stubenrauch et al., 2012). To evaluate the magnitude of this effect, we derived a new GEWEX-like dataset of CA from C-ST v.3 using the same algorithm as in the original GEWEX study. Compared to v.2, CA agree well at high latitudes but are lower by 5-10% over the Tropics for sunlit measurements, (Fig. S4). C-ST v2 and 3 generally remain closer to each other than to C-GOCCP and C-KU.

In line with these results, mapping the standard deviation between the three main CA datasets derived from CALIPSO L1 measurements (C-GOCCP, C-KU and C-ST version 3) shows largest disagreements in tropical subsidence areas and around convection areas (Fig. S5), where low-level fragmented clouds are frequent. These areas show no particular correlation with large passive IDSD (Fig. 1 in the main article). "Large" disagreements are also found in some parts of Antarctica, probably due to misclassifications as clouds of stratospheric clouds and/or endemic atmospheric particles (e.g. blowing snow) occurring at various frequencies in the three algorithms. Note however that these disagreements are relatively small compared to similar passive disagreements. All are documented in the main article (Fig. 3 and 4).

The CALIOP pointing angle was changed in November 2007 to minimize specular reflections from horizontally aligned ice crystals (Noel and Chepfer, 2010; Okamoto et al., 2010), which impact the reliability of optical depth retrievals (Hunt et al., 2009). Such occurrences are rare (less than 0.2% of particulate detections, Liu et al., 2009). The 2007 CA, nonetheless, appears to be ~1% larger than in later years on average. This difference, which a trending study would need to take into account, will have little bearing here.

**S5. Histograms of Scattering Ratios vs. Height.**

Histograms of Scattering Ratio (SR) vs. Height conveniently summarize the properties of a given tropospheric cloud population, describing how its opacity distribution evolves with height. Their production based on CALIOP measurements was introduced in Chepfer et al. (2010), where their construction process from particulate and molecular backscatter is explained in full. In short, bins of SR (lidar Scattering Ratio proportional to the intensity of the lidar signal) were selected to discriminate a large range of atmospheric opacities probed by the CALIPSO lidar, from transparent (SR ~ 1) to very opaque (SR > 50). In clear-sky conditions, SR is typically close to 1, leading to a large representation of such values at all altitude levels in all histograms of this kind. SR > 5 are indicative of cloud presence and used within C-GOCCP to build vertical profiles of cloud fraction. Extremely low values (SR < 0.01) are indicative of total lidar signal attenuation produced by a cloud optical depth larger than 3 to 5 down from the TOA, and occur more frequently as altitude decreases. Counting their occurrences provides a useful metric to evaluate at what altitude the atmosphere is able to scatter the entirety of direct radiation received from above (Chepfer et al., 2014). However, we are not able to discriminate clouds from clear-sky in such totally attenuated signal. This implies that cloud fraction profiles derived from CALIOP (used to create histograms such as those in Fig. 5 and 6) do not include information about a fraction of particularly opaque clouds. An ongoing analysis on our team (Guzman et al., in review) shows large occurrences of totally attenuating clouds at very low altitudes in the Southern Ocean and along the ITCZ. Detections of fully attenuated signals do not appear correlated with large spread among passive CA retrievals.

In addition to SR-Height histograms, we document the annual evolution of the cloud population through monthly averages of vertical cloud fraction (CF) profiles. Part of the C-GOCCP dataset, these were computed by counting occurrences of SR > 5 at each altitude level. A minimum backscatter intensity over molecular was required to avoid false cloud detections at higher altitudes, where backscatter noise fluctuations get close to molecular signal levels. The CF creation process is detailed in Chepfer et al., 2013.

The SR-Height histograms and CF profiles presented here were obtained by considering 2007 C-GOCCP data v2.69. Data analysis (not shown) shows that using later years lead to similar results.

**S6. Relative errors in CF profiles for models**
Figure S8 shows the distribution of errors in CF profiles from models compared to CALIOP observations of CA (as in Fig. 9), but presents the error as relative to the CALIOP CA instead of the absolute difference.

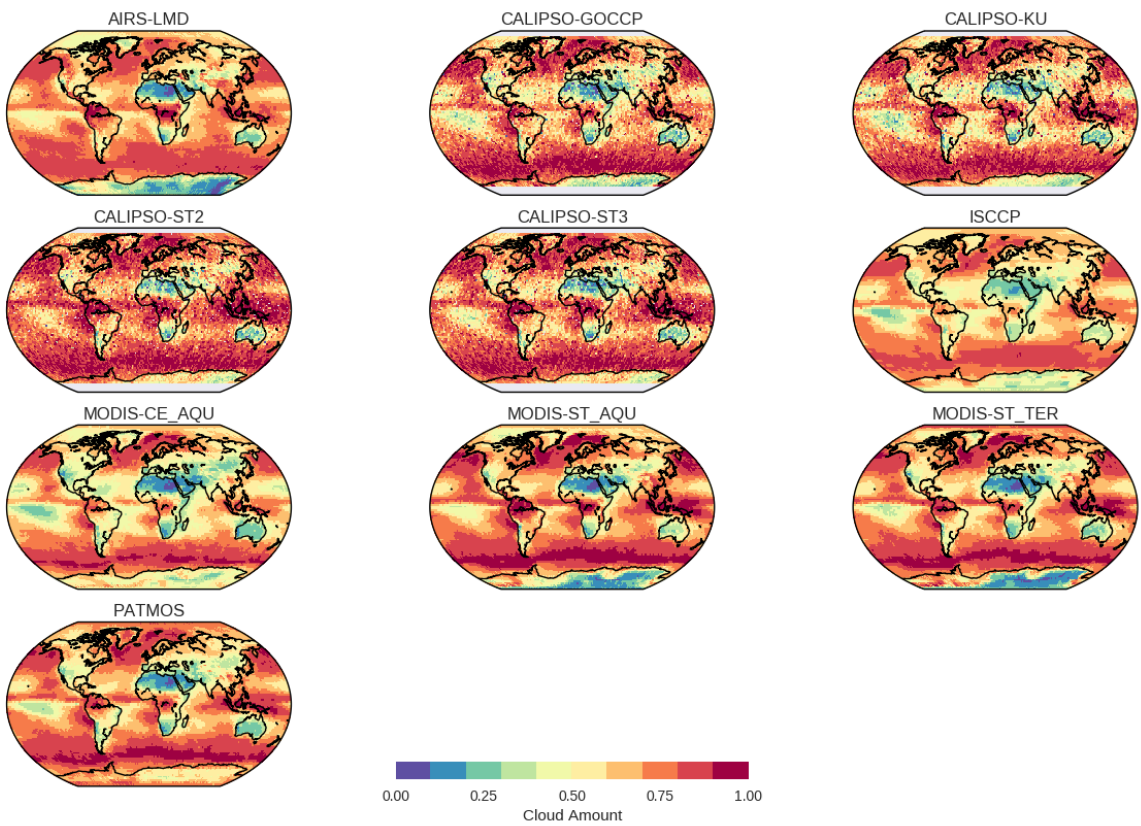

Figure S1. Average AM 2007 CA maps for each passive sensor and the four CALIPSO datasets.

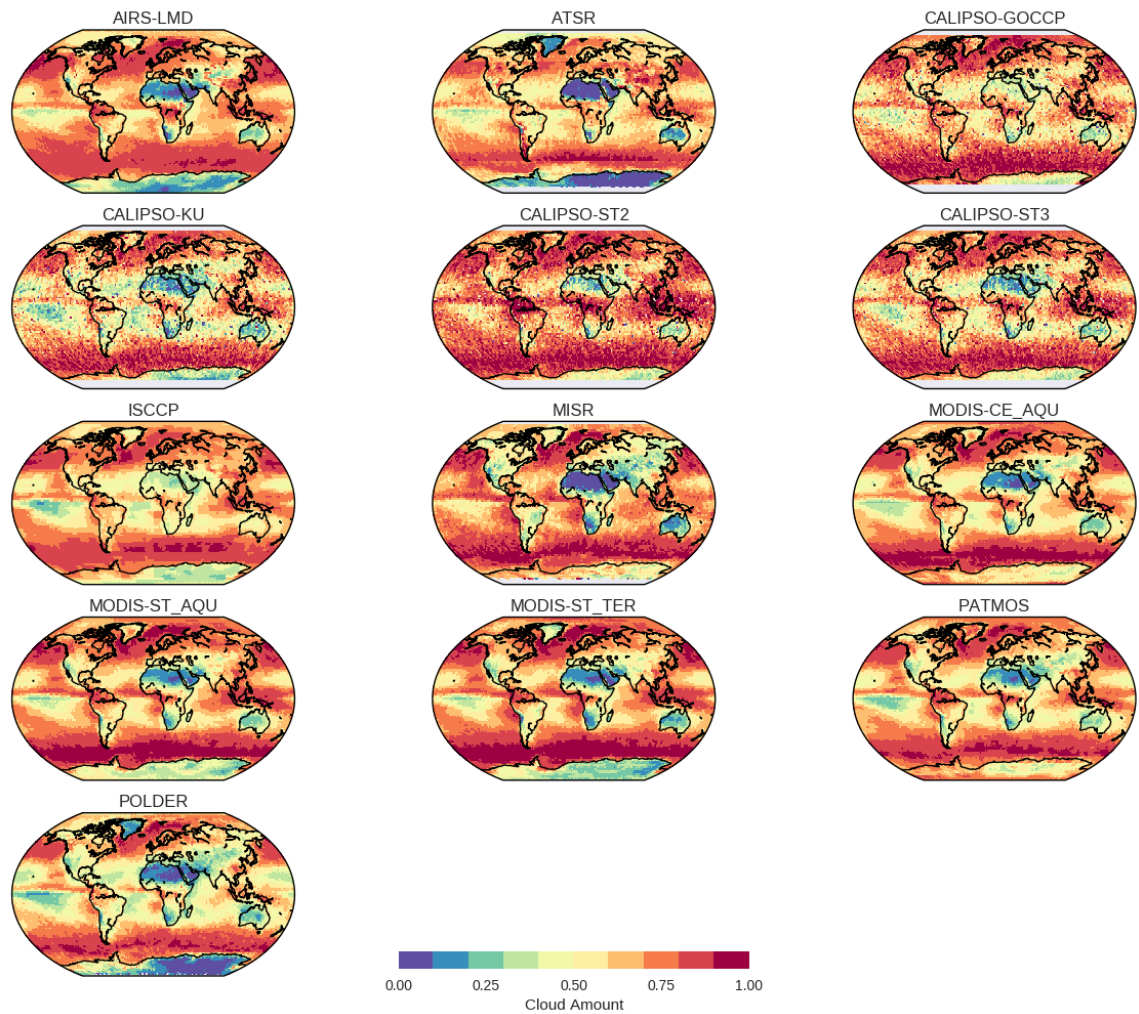

**Figure S2**. Average PM 2007 CA maps for each passive sensor and the four CALIPSO datasets.

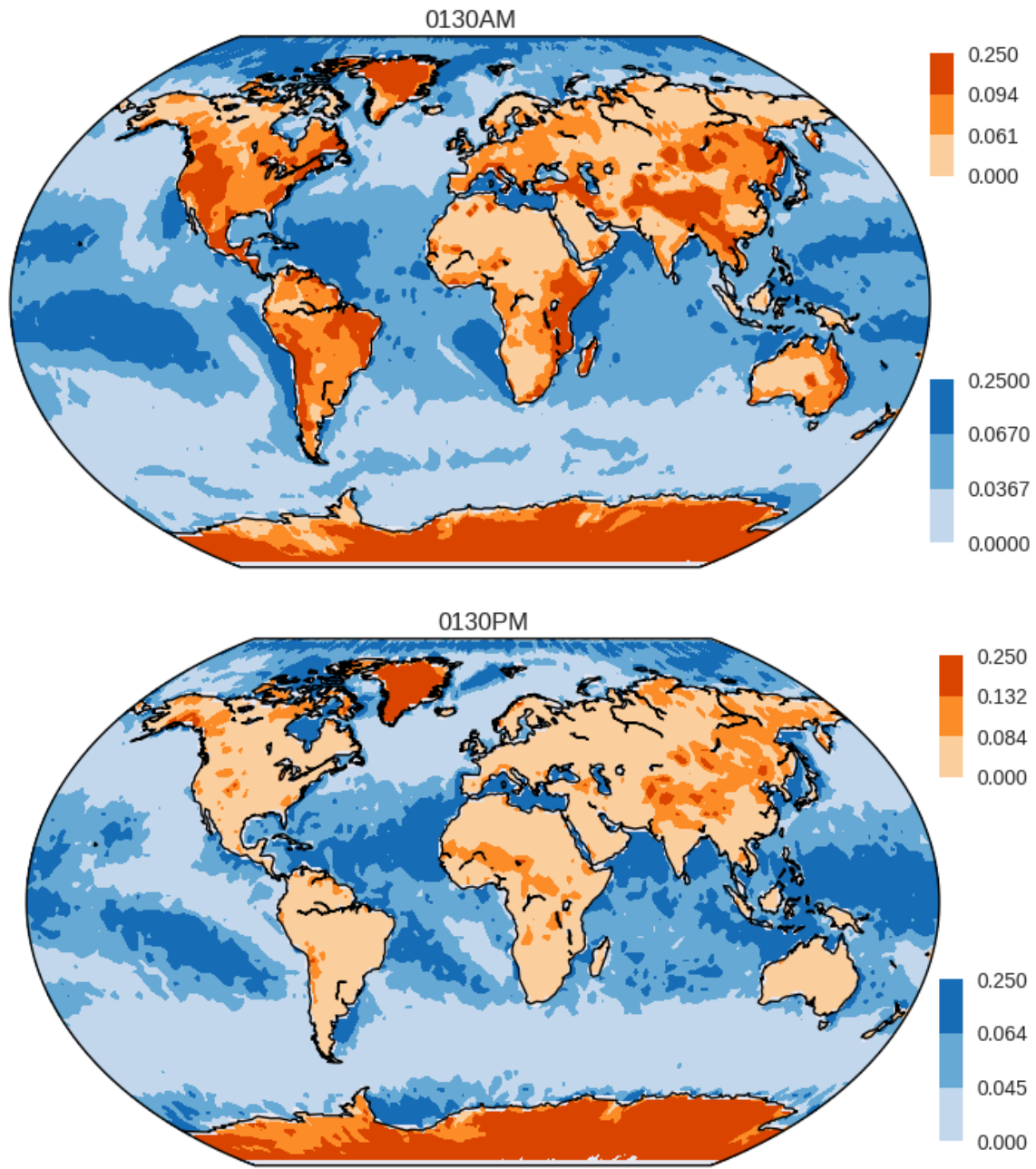

**Figure S3.** Passive Inter-Dataset Standard Deviation for datasets with 0130AM (top) and 0130PM (bottom) local overpass times only

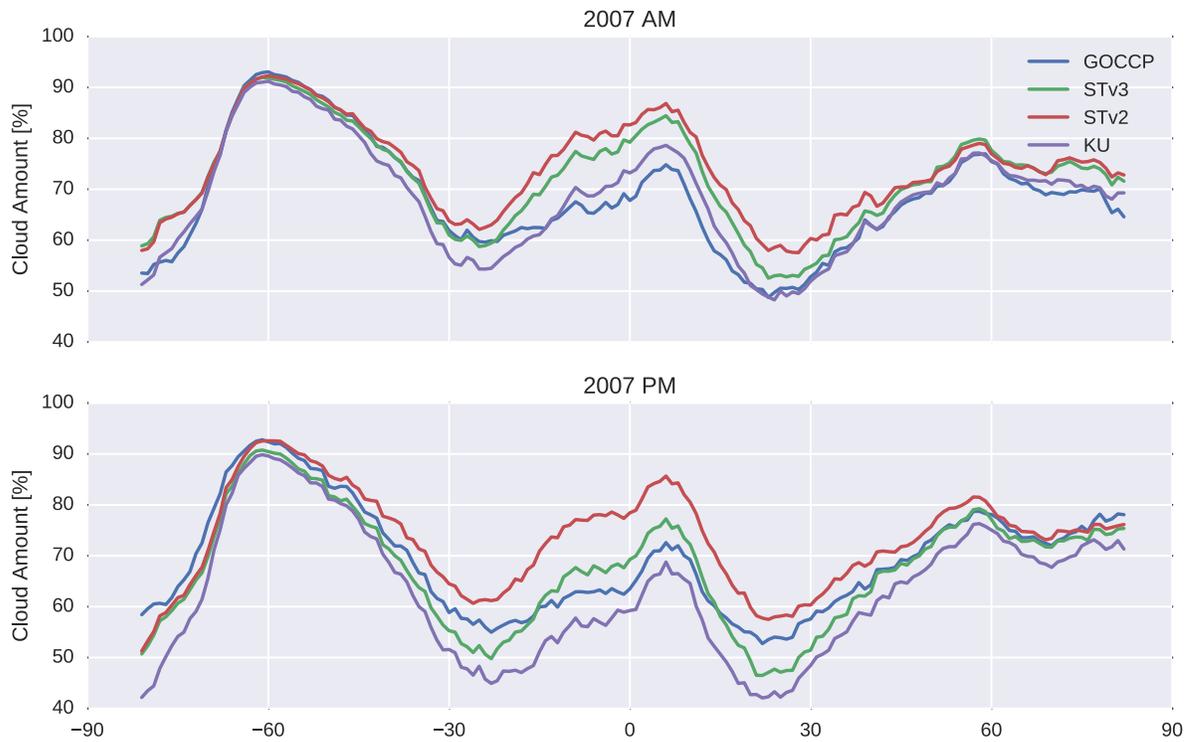

**Figure S4**. Zonal mean of CA reported over 2007 by four datasets derived from CALIOP measurements: CALIPSO-GOCCP, CALIPSO-KU, CALIPSO-ST versions 2 and 3. AM (top) and PM (bottom) retrievals**.**

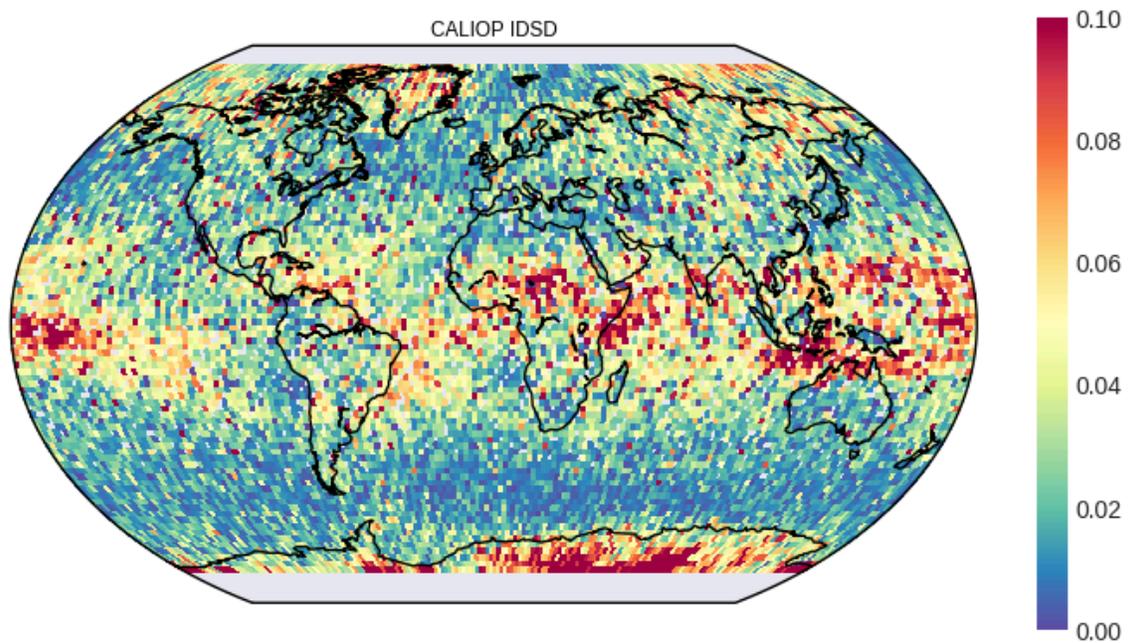

**Figure S5**. Standard deviation between the three annual maps of Cloud Amount derived from CALIPSO-GOCCP, CALIPSO-KU and CALIPSO-ST version 3 AM data.

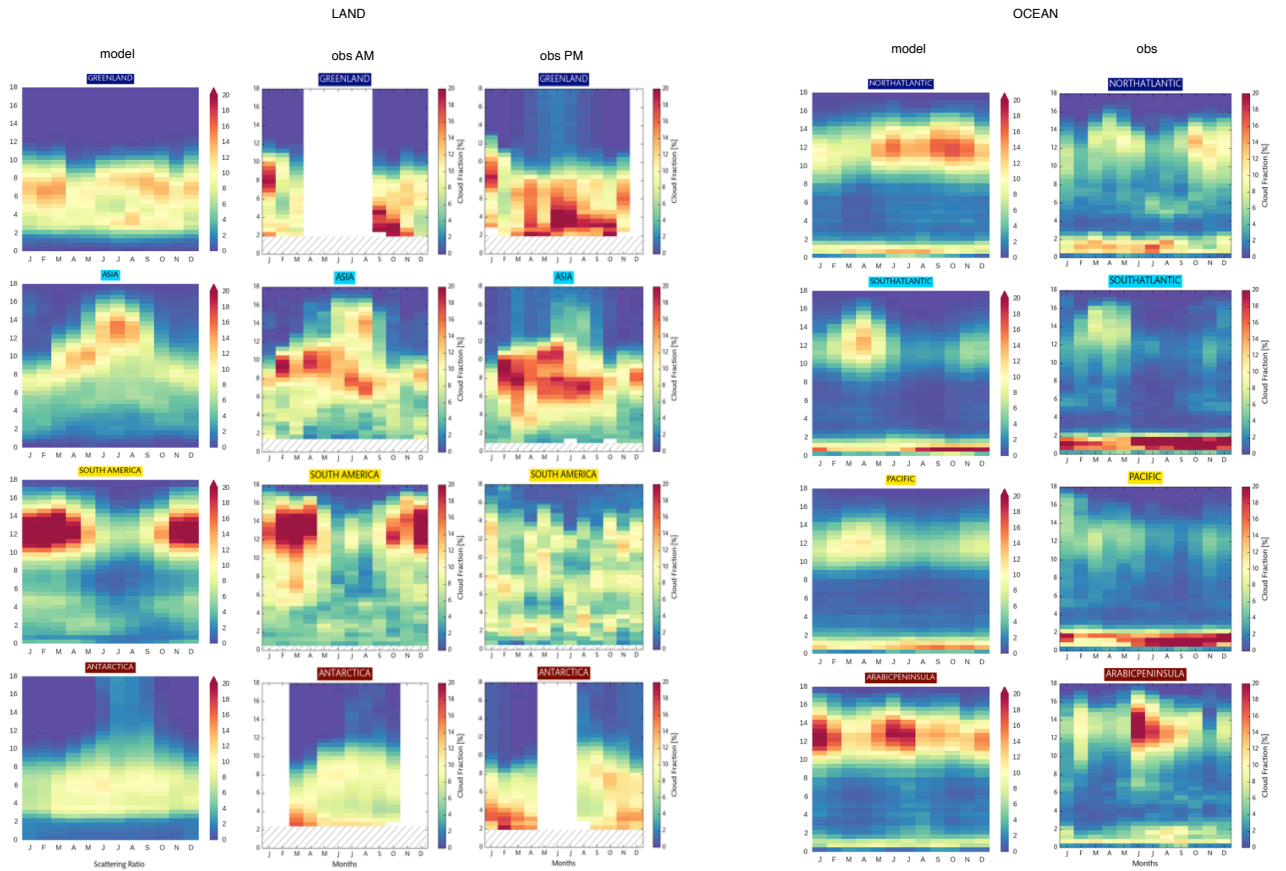

**Figure S6**. Annual cycle of Cloud Fraction profiles within continental regions (left group) and oceanic regions (right group), predicted on average from models (as in Fig. 9) and observed from CALIOP (as in Fig. 5 and 6).

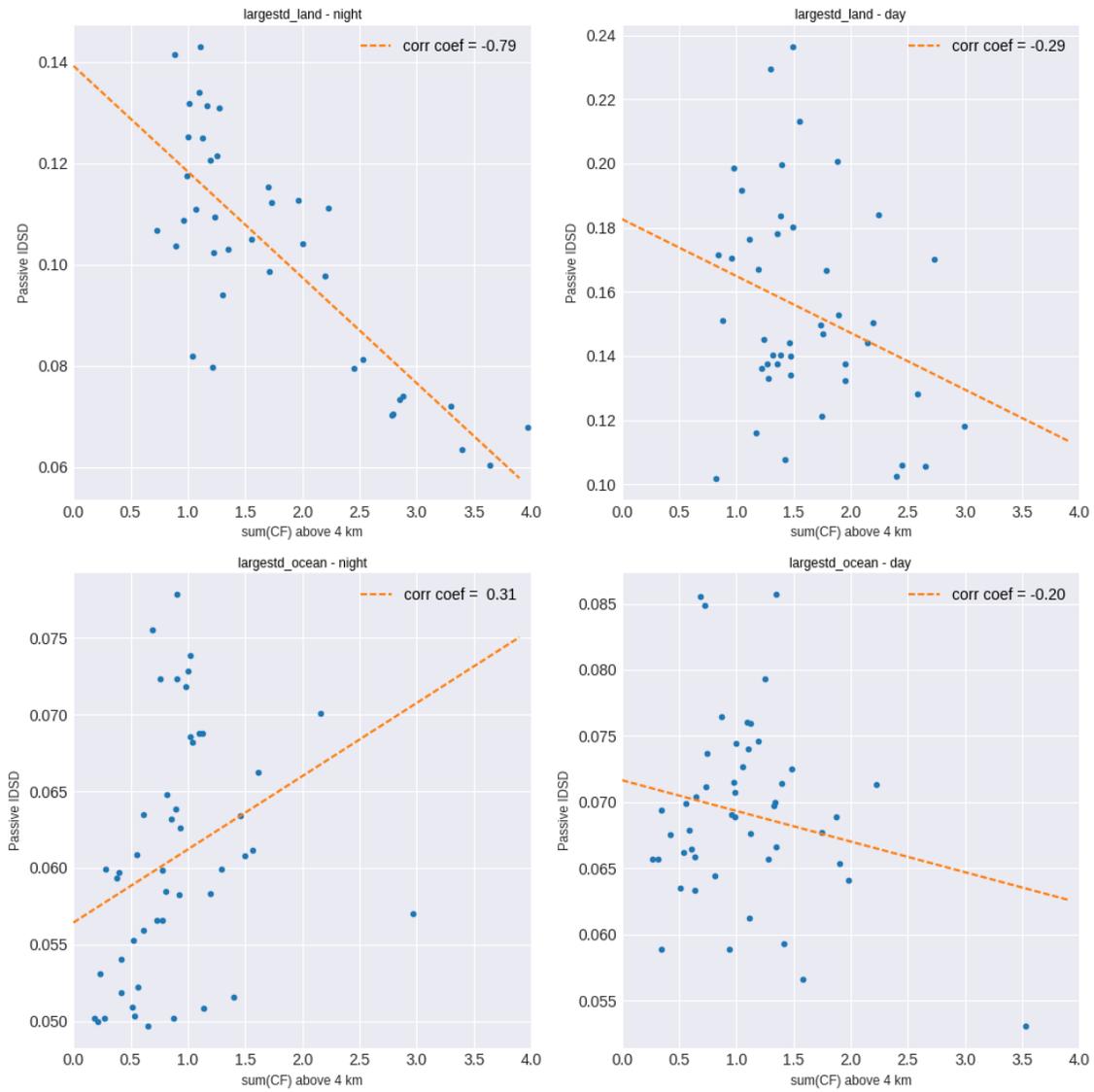

**Figure S7**. Scatterplot of monthly Cloud Fraction profiles shown in Fig 5 and 6 integrated above 4km vs. passive IDSD for all regions.

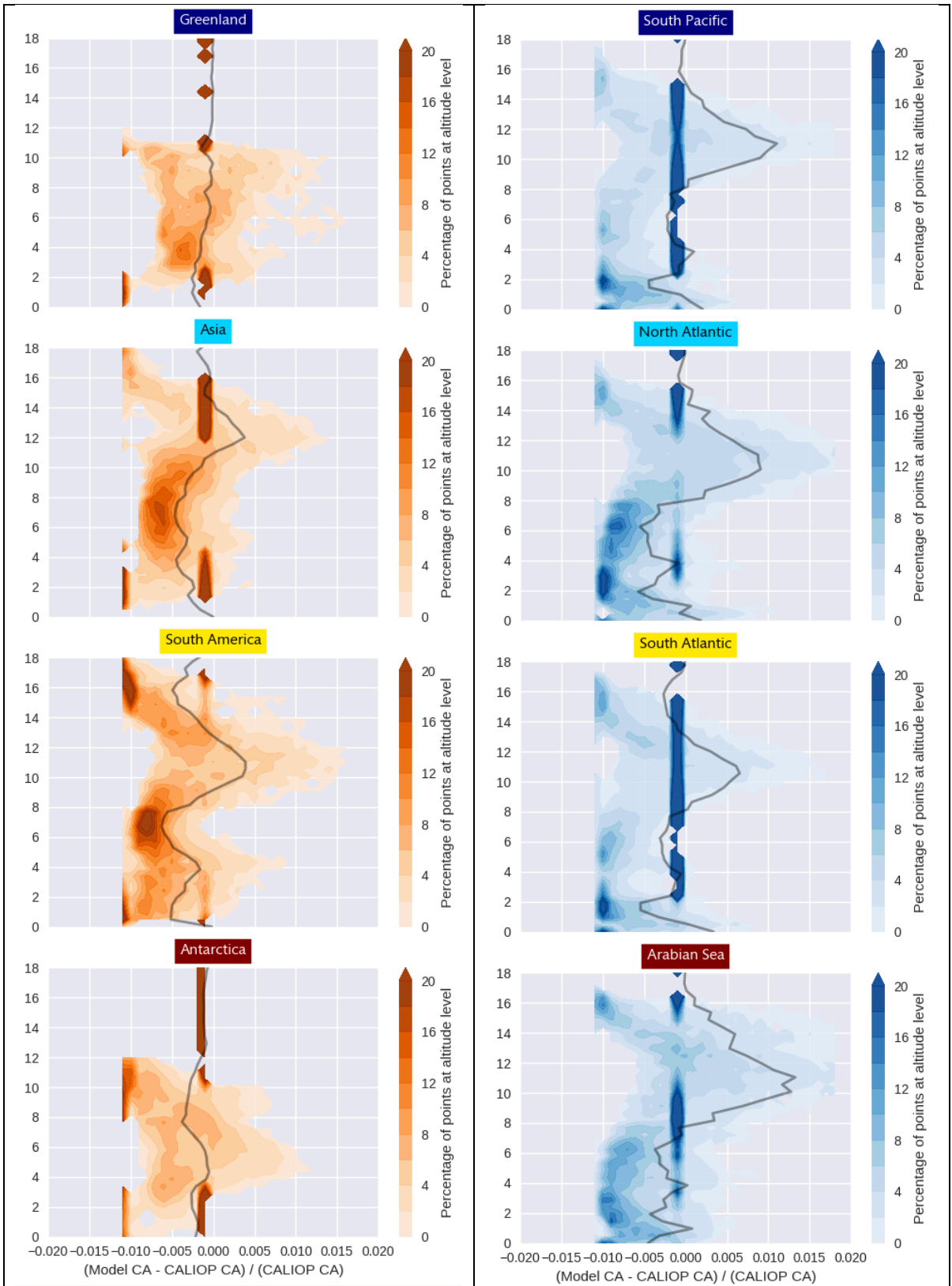

Figure S8: as Fig. 9, but relative to the CA observed from CALIOP.